\documentclass{ima}

\usepackage{graphicx}

\usepackage{amsmath}

\usepackage{natbib}
\usepackage{amssymb}

\usepackage{amsthm}

\usepackage{lineno}
\usepackage{subfig}
\usepackage{algorithm}
\usepackage{algpseudocode}
\usepackage[dvipsnames]{xcolor}
\usepackage[toc,page]{appendix}
\usepackage{mathrsfs}

\newcommand*\patchAmsMathEnvironmentForLineno[1]{%
  \expandafter\let\csname old#1\expandafter\endcsname\csname #1\endcsname
  \expandafter\let\csname oldend#1\expandafter\endcsname\csname end#1\endcsname
  \renewenvironment{#1}%
     {\linenomath\csname old#1\endcsname}%
     {\csname oldend#1\endcsname\endlinenomath}}%
\newcommand*\patchBothAmsMathEnvironmentsForLineno[1]{%
  \patchAmsMathEnvironmentForLineno{#1}%
  \patchAmsMathEnvironmentForLineno{#1*}}%
\AtBeginDocument{%
\patchBothAmsMathEnvironmentsForLineno{equation}%
\patchBothAmsMathEnvironmentsForLineno{align}%
\patchBothAmsMathEnvironmentsForLineno{flalign}%
\patchBothAmsMathEnvironmentsForLineno{alignat}%
\patchBothAmsMathEnvironmentsForLineno{gather}%
\patchBothAmsMathEnvironmentsForLineno{multline}%
}

\usepackage{bbm}
\usepackage{bm}
\usepackage{lineno}
\usepackage{url}
\usepackage{listings}
\definecolor{darkgreen}{rgb}{0.1, 0.14, 0.13}
\usepackage[colorlinks=true,linkcolor=PineGreen,citecolor=PineGreen]{hyperref}

\newcommand{\bbE}{\mathbb{E}}
\newcommand{\bbR}{\mathbb{R}}

\newcommand{\cX}{\mathcal{X}}
\newcommand{\cF}{\mathcal{F}}
\newcommand{\cG}{\mathcal{G}}

\usepackage{color,soul}
\usepackage{mathtools}

\usepackage{gensymb}
\usepackage{array}
\usepackage{multirow}
\usepackage{enumerate}
\newcolumntype{P}[1]{>{\centering\arraybackslash}m{#1}}

\newcolumntype{L}[1]{>{\raggedright\let\newline\\\arraybackslash\hspace{0pt}}m{#1}}
\newcolumntype{C}[1]{>{\centering\let\newline\\\arraybackslash\hspace{0pt}}m{#1}}
\newcolumntype{R}[1]{>{\raggedleft\let\newline\\\arraybackslash\hspace{0pt}}m{#1}}

\usepackage{changes}
\definechangesauthor[name={Reviewer 1}, color = blue]{R1}
\definechangesauthor[name={Reviewer 2}, color = red]{R2}
\definechangesauthor[name={All reviewers}, color = brown]{All}
\definechangesauthor[name={Editor}, color = purple]{Editor}
\definechangesauthor[name={Author}, color = olive]{Author}

\graphicspath{ {./figs/} }

\AtBeginDocument{
  \label{CorrectFirstPageLabel}
  
}

\begin{document}

\title{Learning Stochastic Closures Using Ensemble Kalman Inversion}
\author{
{\sc
Tapio Schneider\thanks{Email: tapio@caltech.edu},
Andrew M. Stuart\thanks{Email: astuart@caltech.edu},
Jin-Long Wu\thanks{Email: jinlong@caltech.edu}} \\[2pt]
California Institute of Technology, Pasadena, CA 91125, USA
}

\maketitle

\begin{abstract}
{Although the governing equations of many systems, when derived from first principles, may be viewed as known, it is often too expensive to numerically simulate all the interactions they describe. Therefore researchers often seek simpler descriptions that
describe complex phenomena without numerically resolving all the interacting components.
Stochastic differential equations (SDEs) arise naturally as models in this context.
The growth in data acquisition, both through experiment and through simulations, provides an opportunity for the systematic derivation of SDE models in many
disciplines. However, inconsistencies between SDEs and real data at short time scales often cause problems, when standard statistical methodology 
is applied to parameter estimation. The incompatibility between SDEs and real data can be addressed by deriving sufficient statistics from the time-series data and learning parameters of SDEs based on these. Here we study sufficient
statistics computed from time-averages, an approach which we demonstrate to lead to sufficient statistics on a variety of problems and which has the secondary 
benefit of obviating the need to match trajectories. Following this approach, we formulate the fitting of SDEs to sufficient statistics from real data as an inverse problem and demonstrate that this inverse problem can be solved by using ensemble Kalman inversion (EKI).
Furthermore, we create a framework for non-parametric learning of drift and diffusion terms
by introducing hierarchical, refinable parameterizations of unknown functions, using Gaussian process regression. We demonstrate the proposed methodology for the fitting of SDE models, first
in a simulation study with a noisy Lorenz '63 model, and then in other
applications, including dimension reduction in deterministic chaotic systems 
arising in the  atmospheric sciences, large-scale pattern modeling in climate dynamics, and simplified
models for key observables arising in molecular dynamics. 
The results confirm that the proposed methodology provides a robust and systematic approach 
to fitting SDE models to real data.}
{Stochastic differential equation; inverse problem; Ensemble Kalman inversion; Gaussian process regression; hierarchical parameterization}
\end{abstract}

\section{Introduction}

\subsection{Overview and Literature Review}

The goal of this paper is to describe a straightforward ensemble-based
methodology that facilitates parameter estimation in ergodic
stochastic differential equations (SDEs), using statistics derived from
time-series data. SDEs arise naturally
as models in many disciplines, and the wish to describe complex phenomena
without explicitly representing all interacting components
within the system make them of widespread interest. Additionally, the
increasing data provides opportunities for the learning 
of stochastic models in previously unforeseen application domains. 
However, SDE models, while often accurate at providing statistical
predictions, may not be compatible with available data at the small scales
where the It\^o calculus model asserts very strong almost sure properties
rarely found in real data; in particular, the quadratic variation 
(variance for a pure Brownian motion) is such
an almost sure property. For this reason, standard statistical methodology 
for parameter estimation in SDEs \citep[see][]{kutoyants2013statistical} is often not suited to fitting models to 
real data. On the other hand, practical experience in the applied sciences
demonstrates the effectiveness of fitting Markovian stochastic processes to sufficient statistics,
typically derived from a long time series by averaging procedures.

Ensemble methods have demonstrable success in the solution of inverse problems, 
based on the use of interacting particle systems driven by the parameter-to-observable 
map; furthermore, they are robust to noisy evaluations of the map \citep{duncan2021}. 
Choosing ergodic averages  as observables and viewing finite-time averages,
or averages over different initializations, as noisy evaluations of the ergodic
average, puts us in a setting where we may apply ensemble methods to effectively 
estimate parameters in SDEs. These ensemble methods are derivative-free,
side-stepping thorny technical and computational issues arising in the parameter-to-observable map for SDEs. They are also inherently parallelizable and scale well to the learning of high-dimensional parameter vectors, making them
a very attractive computational tool.
Finally the formulation of the inverse problem that we 
adopt, using time-averaged data, avoids
the need to determine the latent trajectory variables; the approach
we employ is described in \cite{CES} and \cite{dunbar2020calibration}.

Data-driven methods for extracting simplified models are now
ubiquitous \citep{brunton2019data,coifman2008diffusion,ferguson2011nonlinear,froyland2014computational,klus2018data,giannakis2019data}.
A guiding example that motivates the need for reduced models of
various kinds is the Navier--Stokes equation for fluid motion.
It is often too expensive to numerically simulate all relevant degrees of freedom --- for example turbulence around an aircraft or
convection in Earth's atmosphere. Therefore, stochastic descriptions become important \citep[see, e.g.,][]{majda2012filtering}. In those reduced systems, stochastic models help account for a lack of knowledge and for unresolved variability.
Stochastic models are widespread in applications, including
in biology \citep[see, e.g.,][]{goel2016stochastic,wilkinson2018stochastic}, chemistry
\citep[see, e.g.,][]{leimkuhler2004simulating,tuckerman2010statistical,boninsegna2018sparse},
engineering \citep[see, e.g.,][]{maybeck1982stochastic}, the geophysical
sciences \citep[see, e.g.,][]{majda1999simplified,palmer2001nonlinear,arnold2013stochastic} and the social
sciences \citep[see, e.g.,][]{diekmann2014stochastic}; \cite{gardiner2009stochastic}
provides a methodological overview aimed at applications in the physical and social sciences.

The statistics literature concerning parameter estimation from continuous
time series often starts from the premise that the diffusion coefficient may be 
read off from small increments \citep[see, e.g.,][]{kutoyants2013statistical}, a property that
only holds if the data are consistent with an SDE at arbitrarily fine scales.
The fact that data may be inconsistent with the SDE at small scales
was understood in the finance literature in the early part of this
century, and new models were introduced to address
this incompatibility \citep[see, e.g.,][]{zhang2005tale}.
In subsequent papers \cite[see, e.g.,][]{pavliotis2007parameter,papavasiliou2009maximum}, the setting
of multiscale SDEs was used to elucidate similar phenomena arising from models
in the physical sciences. Subsequent to these works, new methods were introduced
to tackle the inconsistency at small scales, based on subsampling and
other multiscale uses of the data \citep[see, e.g.,][]{zhang2006efficient,papaspiliopoulos2012nonparametric,pokern2009parameter,abdulle2020drift};
see \citet{pavliotis2012parameter} for an overview of some of this work. 
The potential applicability of problems of this type ranges from applications in econometrics, finance and molecular dynamics,
to problems in the
geophysical sciences \citep{cotter2009estimating,kwasniok2009deriving,ying2019bayesian}.

A completely
different approach to circumvent the use of fine-scale properties of the
time series is to compute sufficient statistics from the time-series, and
use these to learn parameters. This approach was recently studied as a systematic
methodology, and then applied, in a
series of papers \citep[see, e.g.,][]{krumscheid2013semiparametric,krumscheid2015data,kalliadasis2015new},
based on statistics computed from multiple realizations and multiple initial conditions;
alternatively one can use a single very long time-series. In the community of modeling climate variability, using time averages as data to estimate a stochastic model has been explored for decades \citep[see, e.g.,][]{hasselmann1976stochastic,frankignoul1977stochastic,penland1993prediction}. Although most of them focused on linear SDEs, the possible extension to nonlinear SDEs was discussed by \citet{hasselmann1988pips}. The use of statistics of time series (e.g., moments, autocorrelation functions or
power spectral densities) for estimating model parameters is common for discrete time series models, such as autoregressive (AR) or autoregressive moving average (ARMA) models \citep[see, e.g.,][]{neumaier2001estimation,lutkepohl2013introduction,brockwell1991time}. Another approach to parameter estimation in dynamical systems is known as state augmentation \citep[see, e.g.,][]{anderson2001ensemble}. 
These methods proceed by augmenting the state to include parameters, and then using
state estimation techniques such as the particle filter \citep[see, e.g.,][]{smith2013sequential}, the unscented Kalman filter
\citep[see, e.g.,][]{julier2000new,albers2017personalized}
and iterative ensemble Kalman smoothers \citep{evensen2019accounting}. However, the augmentation approach 
is notoriously sensitive in many practical settings \citep[see, e.g.,][]{doucet2001introduction} and we do not pursue it here.
 
In this paper, we build on
the approach pioneered in \citet{krumscheid2013semiparametric} and use 
finite-time averaged data. As a consequence, viewing the perfect 
parameter-to-data map as being an ergodic (infinite-time) average, 
we have access to only approximate, noisy evaluations of the desired
parameter-to-data map, computed from finite-time averages. 
For this reason ensemble Kalman methods provide 
a desirable methodology \citep{duncan2021} because they are 
black-box, derivative-free
and robust to noisy parameter-to-data evaluations.
A central question when
summarizing data, as we do when computing time-averages,
is whether or not the summary statistics are sufficient to identify
the parameters of interest; the reader may refer to the literature
on approximate Bayesian computation (ABC) for discussion of this
issue \citep{sisson2018handbook,fearnhead2012constructing}. The
paper by \cite{wood2010statistical} provides an example of related ideas
applied in the context of learning chaotic dynamical systems; as in our
work, and the work \citet{krumscheid2013semiparametric},
it employs summary statistics that eliminate problems arising
from sensitive dependence on initial conditions. Links between
ensemble Kalman methods and ABC are discussed in \cite{nott2012ensemble}.

With the aim of setting our algorithmic approach in
context, we now describe the specific form of ensemble Kalman inversion that
we use in this paper (EKI), its relationship to other ensemble 
Kalman methods that are widely used for  both state and parameter
estimation, and the acronyms used to describe those related algorithms.
The methods were originally
introduced for state estimation using both filtering
(the EnKF, see \cite{evensen1994sequential}) and 
smoothing (the ES, see \cite{van1996data}). 
Ensemble methods are now also used to simultaneously
estimate the trajectory of a dynamical system, and its parameters.
Furthermore, iterating within the ``analysis'' step of the
data assimilation cycle has proven effective 
(see \cite{gu2007iterative,li2007iterative,sakov2012,bocquet2012combining,bocquet2014}); the terminology 
ensemble randomized maximum likelihood (sequential-EnRML),
iterative ensemble Kalman filter (IEnKF) and
iterative ensemble Kalman smoother (IEnKS) is adopted to
describe the algorithms used.
These approaches have recently been used to learn about model
error in dynamical systems, 
or parameters describing the statistics of model error in dynamical
systems \citep{pulido2018,bocquet2020}. However, there are many inverse
problems in which dynamics are not present, or in which state estimation is
not a desired goal of the computation. The papers 
\citep{chen2012ensemble,emerick2013ensemble,evensen2018analysis}
introduced ensemble Kalman methodologies
directly focussed on the solution of general inverse problems, without
reference to state estimation in a dynamical system; 
these methods go by various names,
including the iterative ensemble smoother (IES), 
batch ensemble randomized maximum likelihood
(batch-EnRML) and multiple data-assimilation ensemble smoother (ES–MDA).
In the context of solving inverse problems, 
the methods IES, batch-EnRML and ES–MDA adopt a Bayesian
framework, and iterate over a prescribed
set of iterations, starting from prior samples, with goal being the
generation of approximate posterior samples.
The seminal paper by \cite{reich2011dynamical} made an
important step by connecting ensemble methods
with sequential Monte Carlo for Bayesian inverse problems,
and additionally remarks that 
ensemble Kalman methods could be used for for optimization 
(classical rather than Bayesian inversion) and discusses 
possible stopping criteria.

The ensemble Kalman methods are formulated and evaluated as an 
optimization approach to general 
inverse problems in \citet{iglesias2013ensemble} 
and the idea of incorporating constraints, widely useful in applications, 
is described in \citet{albers2019ensemble}.
\citet{iglesias2013ensemble} demonstrate that, for a variety 
of PDE inverse problems, ensemble Kalman methods implemented without localization perform 
well. These methods may be viewed as minimizing the model-data misfit,
and regularization is introduced through the invariant subspace
property of the algorithm: the minimization is confined to the
space spanned by the initial ensemble. We refer to this 
methodology, and its continuous time analogues
(see \cite{schillings2017analysis}),  
as ensemble Kalman inversion (EKI), recognizing that it is
a variant on the creative ideas developed
in \cite{chen2012ensemble} and \cite{emerick2013ensemble}. We emphasize, however,
that the goal of the EKI approach is to solve an optimization formulation
of the inverse problem, rather than the Bayesian inverse problem
that motivates the approaches in \cite{chen2012ensemble} and \cite{emerick2013ensemble}. 
Overfitting can still be an issue for EKI methods, and
may be addressed by generalizing ideas standard in the classical
optimization approach to inverse problems \cite{engl1996regularization},
as anticipated in \cite{reich2011dynamical};
in particular Levenberg-Marquadt analogues of EKI are
studied in \cite{iglesias2015iterative,iglesias2016regularizing},
and a Tikhonov extension, TEKI, is described in
\cite{chada2020tikhonov}. Bayesian regularization is also
possible, leading to the ensemble Kalman sampler in \cite{garbuno2020interacting}.
In this paper the problems studied
typically have free parameters with dimension smaller  than or
similar to  the dimension of the data-set; thus
overfitting is not an issue and regularization is not employed. 
Other derivative-free optimization methods could
also be employed, such as the consensus-based optimization procedures
described in \cite{carrillo2018analytical}. An important aspect of
the success of the ensemble Kalman methods we use is their
affine invariance, a concept introduced in
\cite{goodman2010ensemble} for Monte Carlo methods; its
significance for ensemble methods was identified in
\cite{garbuno2020affine} and explains the problem-independent
convergence rates obtained by the method, provably in the
case of linear problems \cite{garbuno2020interacting}.

Within the EKI-based
parameter estimation methodology, we employ ideas from Gaussian process regression
(GPR) \citep[see, e.g.,][]{rasmussen2006gaussian} to parameterize unknown functions; this refinable,
hierarchical approach to function representation leads to novel nonparametric methods
for function learning. Our approach builds on preceding, nonhierarchical approaches
to inversion using GPR such as that described in \citet{xiao2016quantifying}. The concept of learning the values at some fixed nodes of a Gaussian process, 
known as the pilot point method~\citep[see, e.g.,][]{doherty2010approaches}, has also been extensively explored by the groundwater modeling community in the context of inverse problems.

\subsection{Our Contributions}

Our contributions in this paper are as follows:

\begin{enumerate}
    
    \item We formulate parameter estimation in ergodic SDEs as a classical inverse problem for the
    parameter-to-data map defined by ergodic averaging.

\item We develop algorithms suited to the setting in which the
parameter-to-data map is available only through 
finite-time averages, which may be viewed as providing noisy approximations
of the ideal ergodic averages.

\item Through a sequence of examples described below we demonstrate that
a simple and straightforward implementation of ensemble Kalman methods, 
the EKI, is well-adapted to solving the inverse problem in which
only noisy approximate evaluations of the parameter-to-data map 
are available.
    
\item Within the EKI we demonstrate the utility of hierarchical parameterizations of unknown functions, using GPR.

\end{enumerate}

We demonstrate the methodology when applied to a variety of examples:

\begin{itemize}
    
    \item A simulation study that employs
    a noisy (SDE) version of the Lorenz '63 model.
    
    \item Reduction of the Lorenz '63 ODE model to a two dimensional SDE
    in coordinates computed by applying PCA to the ODE data.
    
    \item The multiscale Lorenz '96 ODE model, seeking an SDE closure in the
    slow variables alone.
    
    \item To fit a stochastic delay differential equation (SDDE)
    to El Niño–Southern Oscillation data.
    
    \item To fit an SDE that describes fluctuations in the dihedral 
    angle of a butane molecule.

\end{itemize}

In considering these simulation studies and real-data examples, we demonstrate the
effectiveness of the methodology to find parameters, and we evaluate the accuracy of various stochastic models derived from data.
In section \ref{sec:PF}, we formulate the inverse problem of interest,
and introduce four example problems to which we will apply our methodology.
Section \ref{sec:Alg} describes the ensemble Kalman methodology 
we employ to solve the inverse problem, as well as a discussion
of the novel hierarchical
Gaussian process based representation that we employ to 
represent, and learn, unknown functions. In section \ref{sec:N},
we describe numerical results relating to each of the four example
problems. We conclude in section \ref{sec:C}.

\section{Problem Formulation}
\label{sec:PF}

The aim of this work is to estimate parameters $\theta \in \Theta$ in the SDE
\begin{equation}
    \frac{dx}{dt}=f(x;\theta)+\sqrt{\Sigma(x;\theta)}\frac{dW}{dt}
\end{equation}
where $x \in \mathbb{R}^n$,  $f: \mathbb{R}^n \times \Theta \mapsto \mathbb{R}^n$ 
and $\Sigma: \mathbb{R}^n \times \Theta \mapsto \mathbb{R}^{n \times n}.$ For
all of the examples $\Theta \subseteq \mathbb{R}^p$, but the algorithms we use
extend to include the nonparametric setting in which $p=\infty$. 
However, in the specific applications considered here, $p$ 
is small, of $\mathcal{O}(10)$; and
in many other applications envisaged,
$p$ may be considerably smaller than the dimension
$n$ of the state space.
Among several parameterizations used in this paper,
we showcase a GPR based method for hierarchical function representation that is refinable. 

We assume that the SDE is ergodic and let $\bbE$ denote expectation with respect to the
stationary process resulting from this ergodicity. If $x(\cdot;\theta) \in \cX:=C(\bbR^+;\bbR^n)$ denotes 
a solution of the SDE started in a statistical stationary state and $\cF:\cX \mapsto \bbR^q$ is a function on the
space of solution trajectories, where $q$ denotes the dimension 
of data space, then define $\cG:\Theta \mapsto \bbR^q$ by
\[\cG(\theta)=\bbE \cF(x(\cdot;\theta)).\]
We wish to solve the inverse problem of 
finding $\theta$ from noisy 
approximate evaluations of $\cG(\theta)$.
The noisy approximate evaluations arise from the fact that we will use finite-length trajectories to
approximate the expectation defining $\cG(\theta)$; averages over initial conditions and/or
realizations of the noise could also be used. The following remark highlights the
various observables $\cF$ that we will use in this paper.

\begin{remark}
We will use  $m^\textrm{th}-$moments of vector $x$ at time $t=0$:
\[\cF_m(x(\cdot))=\Pi_{j \in M}x_j(0),\]
where $x_j$ denotes the $j^\textrm{th}$ element of vector $x$, and $M$ is a subset of cardinality $m$ comprising indices (repetition allowed) from
$\{1,\cdots, n\}$, leading to the ergodic average $\cG_m$.
We will also use $\cF_{ac}(x(\cdot))=x(t) \otimes x(0)$, leading through ergodic averaging
to the auto-correlation function $\cG_{ac}$ of the stationary
process. And finally we will use
$\cG_{psd}$ to denote parameters of a polynomial fit to the logarithm of the
power spectral density; recall that the power spectral density is the Fourier transform
of the auto-correlation function of the stationary process.
All of the functions $\cG_m, \cG_{ac}$ and $\cG_{psd}$ 
can be approximated by time-averaging.

It is instructive to think of $\cG(\theta)$ as the 
infinite-time average of the quantities of interest so that, assuming 
ergodicity, the dependence of the initial condition of the trajectory
generating the data disappears. By doing so, we obtain an inverse 
problem in which the latent variable, the trajectory itself, disappears
from the inference problem. This is distinct from many other approaches
to parameter estimation in which trajectories and parameters are
jointly inferred \citep{pulido2018,bocquet2020}. In practice, we have only finite time
averages available, which means that $\cG(\theta)$ is only available to
us through approximate, noisy evaluations, with noise entering through the
dependence on the initial condition and through sensitive dependence on
initial conditions. The ensemble methods described in the
next section are demonstrably and provably
effective in dealing with the setting in which only approximate, noisy 
evaluations of $\cG(\theta)$ are available \citep{duncan2021}. 
\end{remark}

We now describe four examples that will be used to illustrate the methodology.

\begin{example}[Lorenz 63 System]
\label{ssec:L63}
The Lorenz equations \citep[see, e.g.,][]{lorenz1963deterministic} are a system of three ordinary differential equations
taking the form
\begin{equation}
\label{eq:l63}
    \dot{x} = f(x),
\end{equation}
where $x=[x_1,x_2,x_3]^\top$ and $f:\mathbb{R}^3 \mapsto \mathbb{R}^3$ is given by
\begin{equation}
\label{eq:fl63}
\begin{aligned}
f_1(x)&=\alpha(x_2-x_1), \\
f_2(x)&=x_1(\rho-x_3)-x_2, \\
f_3(x)&=x_1x_2-\beta x_3.
\end{aligned}
\end{equation}
We are interested in the noisy version of these equations, in the form
\begin{equation}
\label{eq:l63n}
    \dot{x} = f(x)+\sqrt{\sigma}\dot{W}.
\end{equation}
This SDE will be used in a simulation study to illustrate our methodology.

We will also use the Lorenz 63 model \eqref{eq:l63}, \eqref{eq:fl63} written
in a new coordinate system computed by means of PCA, as introduced in
\citet{selten1995efficient} and \citet{palmer2001nonlinear}. This amounts to introducing
new coordinates $a=Ax$, in which we obtain
\begin{equation}
\label{eq:l63pca}
    \dot{a} = g(a),
\end{equation}
where $x=[x_1,x_2,x_3]^\top$ and $g:\mathbb{R}^3 \mapsto \mathbb{R}^3$ is given by
\begin{equation}
\label{eq:gl63}
\begin{aligned}
g_1(a)&=2.3a_1-6.2a_3-0.49a_1a_2-0.57a_2a_3, \\
g_2(a)&=-62-2.7a_2+0.49a_1^2-0.49a_3^2+0.14a_1a_3, \\
g_3(a)&=-0.63a_1-13a_3+0.43a_1a_2+0.49a_2a_3.
\end{aligned}
\end{equation}
The coordinate $a_3$ contains around $4\%$ of the total variance of the system. In
\citet{palmer2001nonlinear}, this was used as an argument to seek a stochastic dimension
reduction of the model, in discrete time, in the variables $a_1, a_2$ alone.
We reinterpret this idea in continuous time
and use our methodology to evaluate the idea, using data from \eqref{eq:l63pca}, \eqref{eq:gl63}
to study the fidelity possible when fitting an SDE of the form
\begin{equation}
\label{eq:l63pcat}
\begin{aligned}
\dot{a}_1&=2.3a_1-0.49a_1a_2+\psi_1(a_2)+\sqrt{\sigma_1(a_2)}\dot{W}, \\
\dot{a}_2&=-62-2.7a_2+0.49a_1^2+\psi_2(a_1)+\sqrt{\sigma_2(a_1)}\dot{W}, \\
\end{aligned}
\end{equation}
where the functions $\psi_1(\cdot)$, $\psi_2(\cdot)$, $\sigma_1(\cdot)$, $\sigma_2(\cdot)$ are represented by Gaussian process regression as described in detail subsection~\ref{ssec:GPR}. 

\end{example}

\begin{example}[Lorenz 96 System]
\label{ssec:L96}
The Lorenz 96 multiscale system \citep[see, e.g.,][]{lorenz1996predictability} describes the evolution of two sets of variables, denoted by
$x_k$ (slow variables) and $y_{j,k}$ (fast variables):
\begin{equation}
\label{eq:l96}
    \begin{aligned}
    \frac{dx_k}{dt}&=-x_{k-1}(x_{k-2}-x_{k+1})-x_k+F-hc\overline{y}_k, \quad k \in \{1,\dots, K\}, \\
    \frac{1}{c}\frac{dy_{j,k}}{dt}&=-by_{j+1,k}(y_{j+2,k}-y_{j-1,k})-y_{j,k}+\frac{h}{J}x_k, \quad (j,k) \in \{1,\dots, J\} \times \{1,\dots, K\}\\
    &\quad x_{k+K}   = x_k, \quad y_{j,k+K} = y_{j,k}, \quad y_{j+J,k} = y_{j,k+1}.
    \end{aligned}
\end{equation}
The coupling term $hc\overline{y_k}$ describes the impact of fast dynamics on the slow dynamics,
with $\overline{y}_k$ being the average
\begin{equation}
\overline{y}_k=\frac{1}{J}\sum_{j=1}^J y_{j,k}.
\end{equation}
We work with the parameter choices as in \citet{schneider2017earth}, which are the same parameter choices as in \citet{lorenz1996predictability}. 
Specifically, we choose $K=36$, $J=10$, $h=1$, and $F=c=b=10$. By assuming a spatially
homogeneous (with respect to $j$) equilibrium in the fast dynamics, fixing $k$ and $x_k$ as in \citet{fatkullin2004computational}, we obtain
\begin{equation}
\overline{y}_k=\frac{h}{J}x_k.
\end{equation}
We will seek a stochastic closure for the slow variables $\{x_k\}$
encapsulating systematic deviations from,
and random fluctuations around, this simple balance:
\begin{equation}
\label{eq:l96c}
\begin{aligned}
\dot{X_k}&=-X_{k-1}(X_{k-2}-X_{k+1})-X_k+F-\frac{h^2c}{J}X_k + \psi(X_k) + \sqrt{\sigma}\dot{W},\\
X_{k+K}&=X_k.
\end{aligned}
\end{equation}
Using data from \eqref{eq:l96}, we will fit a parameterized $\psi(\cdot)$ and the
noise level $\sigma.$ 
\end{example}

\begin{example}[El Ni\~no–Southern Oscillation]
\label{ssec:ENSO}

The El Ni\~no–Southern Oscillation (ENSO) \citep[see, e.g.,][]{ENSO} is a well-documented phenomenon, which describes irregularly recurring changes in central and eastern tropical Pacific Ocean temperatures. Figure~\ref{fig:ENSO-data} (time-series and histogram) shows Pacific sea surface temperature (SST) data (mean value has been removed) for years 1870 to 2019. The temperature data are averaged within 5S-5N and 170-120W, a region known as Ni\~no 3.4. It is 
postulated that the mechanism of ENSO can be illustrated by a delayed oscillator model with two time delays determined by properties of the dynamics of the central tropical Pacific Ocean. Specifically, the two time delays are often interpreted as associated with an eastward traveling Kelvin wave and a westward traveling Rossby wave (which also becomes an eastward traveling Kelvin wave after being reflected by the western coastline)~\citep[see, e.g.,][]{tziperman1994nino}. Thus, the time delays can be viewed as known and estimated quantitatively
from the corresponding wave speeds.

\begin{figure}[!htbp]
  \centering
  \subfloat[Time series]{\includegraphics[height=0.25\textwidth]{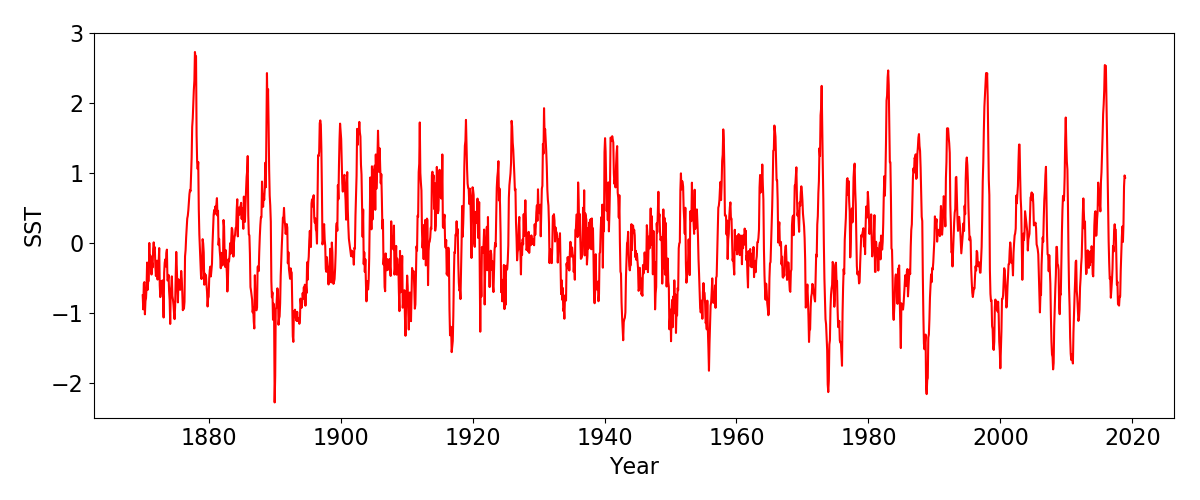}}
  \subfloat[Histogram]{\includegraphics[height=0.25\textwidth]{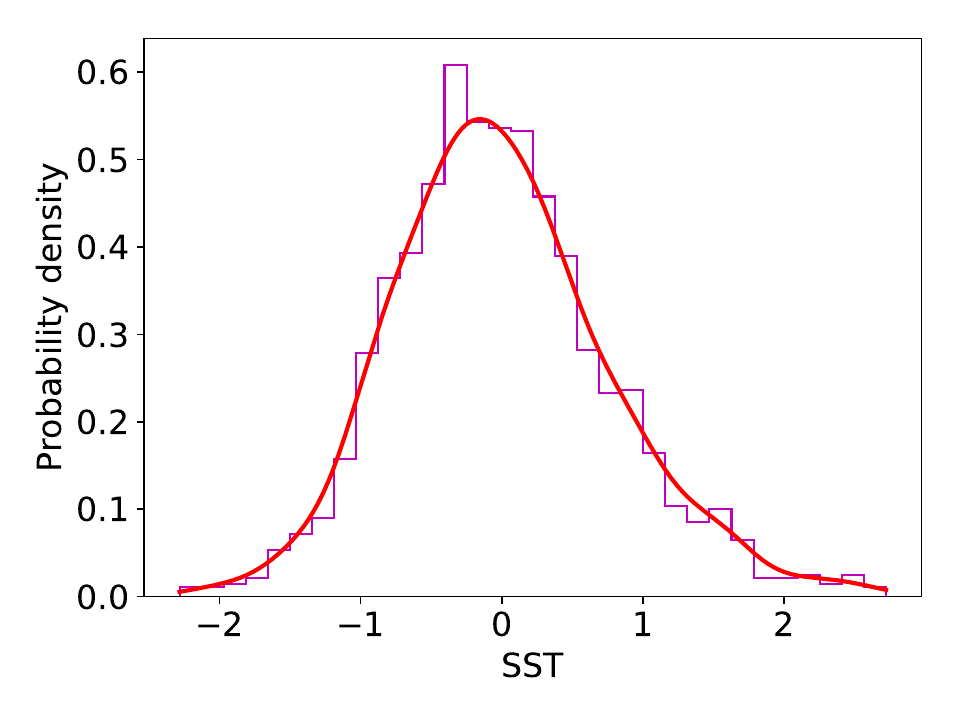}}
    \caption{Illustration of ENSO data \citep{ENSO} from year 1870 to 2019 (the time interval between two adjacent data points is one month), where SST stands for sea surface temperature. It should be noted that although the left-hand shows a time-series, only time-averaged statistics are used as training data in this work.}
  \label{fig:ENSO-data}
\end{figure}

The aim is to fit a stochastic delay differential equation (SDDE) \citep[see, e.g.,][]{buckwar2000introduction,erneux2009applied} to the ENSO 
data shown in Fig.~\ref{fig:ENSO-data} (time-series and histogram). To be concrete, we will fit a model of the following form \citep{tziperman1998locking}:
\begin{equation}
\label{eq:delay-oscillator}
    \frac{dx}{dt}(t)=a\tanh(x(t-\tau_1))-b\tanh(x(t-\tau_2))-cx(t)+\sqrt{\sigma}\frac{dW}{dt}(t).
\end{equation}
Here delay $\tau_1$ represents the effect of the eastward traveling Kelvin wave, and delay $\tau_2$ represents the effect of the westward traveling Rossby wave.
Since good estimates for $\tau_1$ (1.15 months) and $\tau_2$ (5.75 months) exist, we will simply fit the
four parameters $a,b,c,\sigma$ to time-averaged data computed from the time-series shown in the left-hand panel in Fig.~\ref{fig:ENSO-data}.

\end{example}

\begin{example}[Butane Molecule Dihedral Angle]
\label{ssec:butane}

In many problems arising in molecular dynamics, it is of interest to determine reduced
models describing the behavior of certain functionals derived from the molecular
conformation. An example of such a functional is the dihedral angle in a simple
model of a butane molecule \citep[see, e.g.,][]{schlick2010molecular}. Figure~\ref{fig:dihedral-angle} shows how the dihedral angle is
defined from the four carbon atoms that comprise the butane molecule. 
Figure~\ref{fig:Butane-data} shows the time series and histogram
for this angle, derived from a molecular dynamics model which we now describe. The model comprises a system of twelve
second order SDEs for the $4$ atomic positions in $\bbR^3$, of Langevin type:
\begin{equation}
    \label{eq:Langevin1}
m_0\frac{d^2x}{dt^2}+\gamma_0 \frac{dx}{dt}+\nabla V(x)=\sqrt{\frac{2\gamma_0}{\beta_0}}\frac{dW}{dt}.
\end{equation}
The potential $V$, mass $m_0$, damping $\gamma_0$ and inverse temperature $\beta_0$ characterize the molecule. From the time series of $x \in C(\bbR^+;\bbR^{12})$ generated by this model we fit 
a simplified model for the the dihedral angle. This simplified model 
for $\phi \in C(\bbR^+;\bbR)$ takes the form
\begin{equation}
    \label{eq:Langevin2}
\frac{d^2\phi}{dt^2}+\gamma(\phi) \frac{d\phi}{dt}+\nabla \Psi(\phi)
=\sqrt{2\sigma \gamma(\phi)}\frac{dW}{dt}.
\end{equation}
We will fit parameterized versions of $\gamma$ and $\Psi$ to data for $\phi$ generated
by studying the time-series for the dihedral angle defined by $x$ from \eqref{eq:Langevin1}. Related work, fitting SDE models for the dihedral
angle, may be found in \citet{papaspiliopoulos2012nonparametric} and \citet{pokern2009parameter}. For
a broader introduction to the subject of finding Markov models
for bimolecular dynamics see \citet{djurdjevac2010markov,ferguson2011nonlinear,zhang2017effective,schutte2013metastability}.

\begin{figure}[!htbp]
  \centering
  \includegraphics[width=0.4\textwidth]{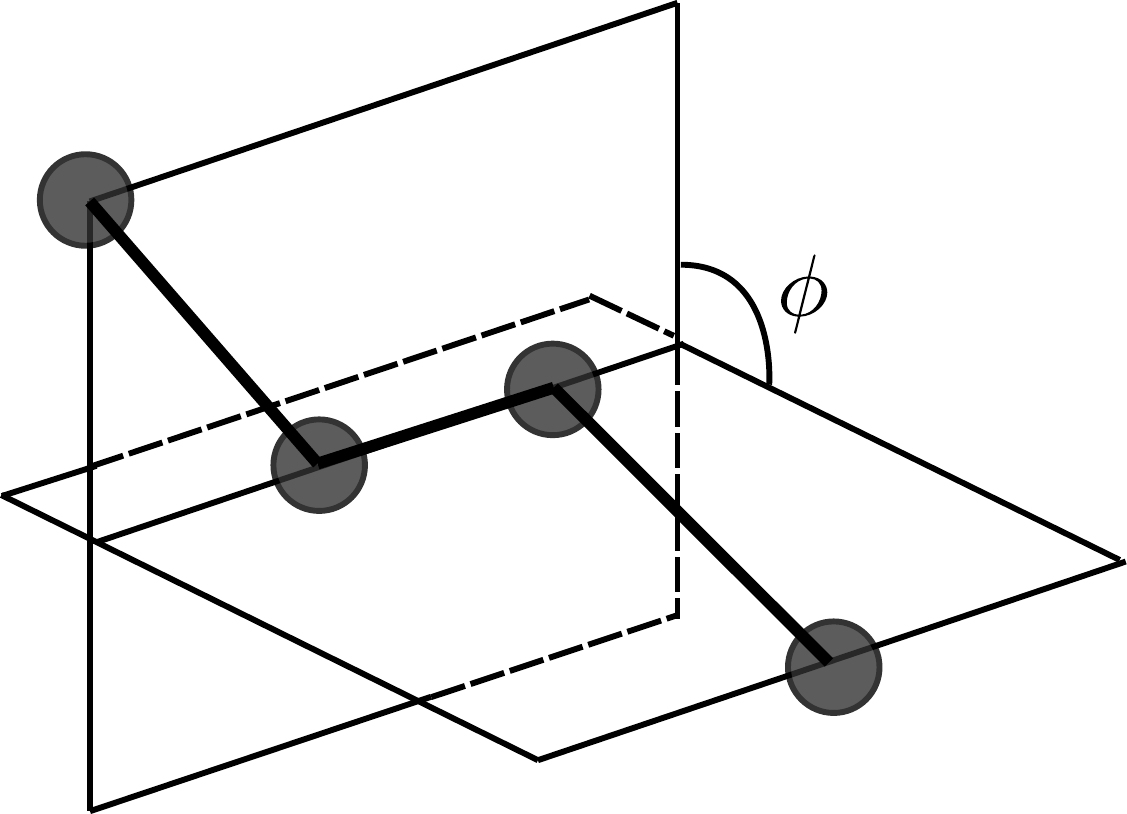}
    \caption{The definition of butane molecule dihedral angle.}
  \label{fig:dihedral-angle}
\end{figure}

\begin{figure}[!htbp]
  \centering
  \subfloat[Time series]{\includegraphics[height=0.25\textwidth]{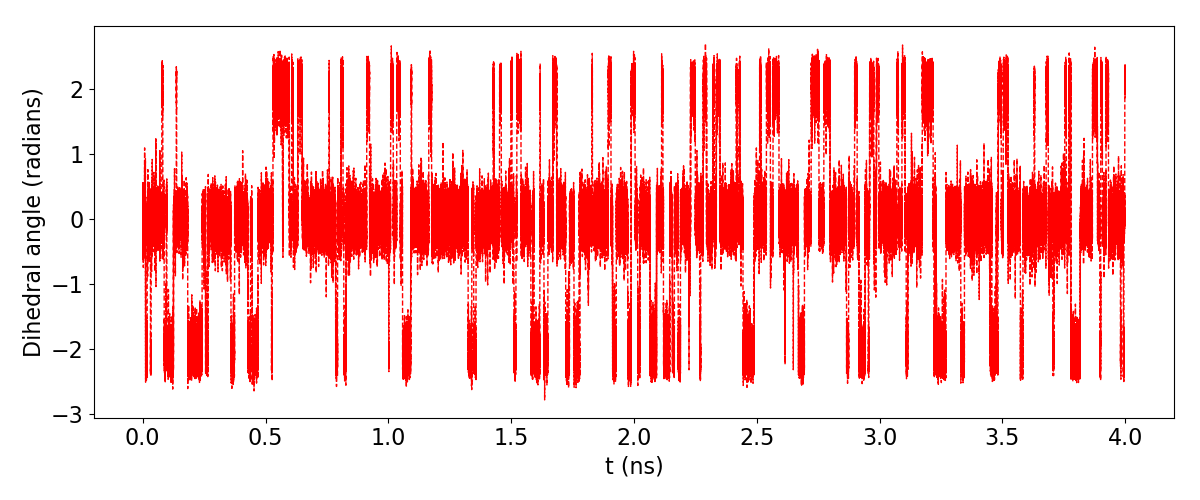}}
  \subfloat[Histogram]{\includegraphics[height=0.25\textwidth]{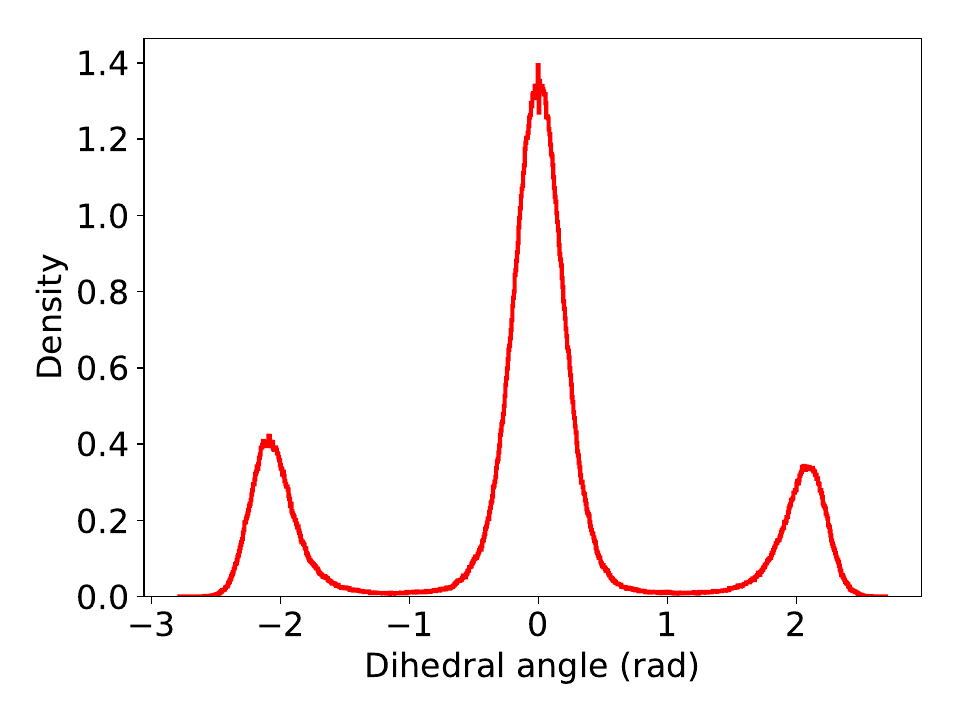}}
    \caption{Illustration of the true butane dihedral angle data (the time interval between two adjacent data points is $10^{-6}$ ns).}
  \label{fig:Butane-data}
\end{figure}

\end{example}

\section{Algorithms}
\label{sec:Alg}

In subsection \ref{ssec:EKI}, we describe the ensemble-based 
derivative-free optimization method that we use to fit parameters.
The approach is based on the algorithms pioneered
in \cite{chen2012ensemble} and \cite{emerick2013ensemble} but is aimed at solving
an optimization formulation of the parameter estimation problem, rather
than a Bayesian formulation; we refer to the specific
iterated form of the algorithm used here as an ensemble Kalman inversion
(EKI) algorithm. In subsection \ref{ssec:GPR}, we discuss
how we use GPR to design hierarchical, refinable parameterizations of unknown functions that we
wish to learn from data.

\subsection{Ensemble Kalman Inversion (EKI)}
\label{ssec:EKI}
Recall that we view the data 
that we are given, $y \in \bbR^J$, as
a noisy evaluation of the function $\cG(\theta)$ defined
by ergodic averages. We use the notation $G_{\tau}(\theta;x_0)$ 
to denote this noisy
evaluation, which arises 
from finite-time averaging of duration $\tau$, started at initial
condition $x_0$; this finite-time averaging approximates the 
ergodic average ($\tau=\infty$) in which dependence on $x_0$ disappears.  
Appealing to central limit theorem results
that quantify rates of convergence towards ergodic averages,
the inverse problem can be formulated 
as follows: given $y \in \bbR^J$, find $\theta \in \Theta$ so that
\begin{equation}
\label{eq:IP0}
y = G_\tau(\theta;x_0) \approx \cG(\theta)+\eta, \quad \eta \sim N\bigl(0,\Gamma(\theta)\bigr).
\end{equation}
Although this central limit theorem argument
leads to a $\Gamma$ which is $\theta-$dependent, in practice we make 
the approximation that it is constant and solve the resulting
Bayesian inverse problem
\begin{equation}
\label{eq:IP}
y=\cG(\theta)+\eta, \quad \eta \sim N\bigl(0,\Gamma\bigr).
\end{equation} 
In this formulation, the trajectory initial condition
$x_0$ disappears from the inference problem. However, $\cG(\theta)$
is not computable and only available to us through noisy approximate
evaluations. We address this issue after introducing the EKI algorithm below.
Estimating $\Gamma$ may be achieved by using time-series 
of different lengths, as explained in \citet{CES} in the specific case of 
$\cG$ derived from moments, $\cF_m.$

The natural objective function associated with the inverse
problem \eqref{eq:IP} is
\begin{equation}
    \label{eq:ob}
    \frac12\bigl\|\Gamma^{-\frac12}\bigr(y-\cG(\theta)\bigr)\bigr\|^2.
\end{equation}
The two primary reasons for using EKI to find approximate minimizers
of this objective function are: (a) it does not require derivatives, which 
can be difficult to compute for SDEs; (b) it is robust to noisy evaluations of the forward map.
The method behaves (provably in the linear case, approximately
in the nonlinear case) like a projected gradient descent
\citep{schillings2017analysis,garbuno2020interacting}, despite not
computing derivatives, and the use of differences promotes  
a desirable averaging making it robust to noisy evaluations of 
the forward map \citep{duncan2021}.

The ensemble Kalman inversion (EKI) algorithm we employ is described
in \citet{iglesias2013ensemble} and \citet{albers2019ensemble}. The algorithm propagates a 
set of $J$ parameter estimates $\{\theta_n^{(j)}\}_{j=1}^J$ through algorithmic 
time $n$, using the update formula
\begin{equation}
\label{eq:EKI}
\theta_{n+1}^{(j)}=\theta_{n}^{(j)}+C_n^{\theta G}\left(C_n^{GG}+\Gamma \right)^{-1}\left(y-\cG(\theta_n^{(j)}) \right).
\end{equation}
The matrix $C_n^{GG}$ is the empirical covariance of $\{\cG(\theta_n^{(j)})\}_{j=1}^J$,
while matrix $C_n^{\theta G}$ is the empirical cross-covariance of 
$\{\theta_n^{(j)}\}_{j=1}^J$ with $\{\cG(\theta_n^{(j)})\}_{j=1}^J$.

We implement the preceding algorithm with two modifications.
The first reflects the benefits that can accrue from randomizing
the data, helping to expand the search through parameter space,
analogous to stochastic gradient descent
\citep{goodfellow2016deep}.  The second reflects the fact that  
the forward model $\cG$ represents ergodic averages, but 
is only available to us approximately through finite time
averages which depend on initial condition.
Thus the algorithm we implement has form
\begin{equation}
\label{eq:EKI2}
\theta_{n+1}^{(j)}=\theta_{n}^{(j)}+C_n^{\theta G}\left(C_n^{GG}+\Gamma \right)^{-1}\left(y_n^{(j)}-G_n^{(j)}) \right).
\end{equation}

We now detail exactly 
how $y_n^{(j)}$ and $G_n^{(j)}$ are defined.
In the experiments reported in this paper,
we add i.i.d. (w.r.t. $j$ and $n$) random mean zero Gaussian
noise with covariance $\Gamma$ to the data $y$ to obtain $y_n^{(j)}$;
however, we have verified that near identical
parameter estimates are obtained in the same number of iterations,
without adding any random noise to $y$, for several of the experiments
reported. Randomization should be interpreted in
the same way that stochastic gradient descent is beneficial in the
optimization of neural networks \citep{goodfellow2016deep}; this is distinct
from its use in the work of \citet{chen2012ensemble,emerick2013ensemble}
where it is used to facilitate posterior sampling, not
optimization. Regarding the forward model we set
$G_n^{(j)}(\cdot)=G_{T}(\cdot; x_0^{(j,n)})$. 
This approximate evaluation of $\cG$,
$G_n^{(j)}$, is found from choosing the initial condition
$x_0^{(j,n)}$ 
at random and i.i.d with respect to both 
ensemble member $j$ and iteration step $n$.
Details are given for each example in what follows. 
Note that the finite time $T$ in $G_n^{(j)}$ is typically
different from $\tau$ arising in the time-averaged data.

Here we use the algorithm in settings where the number 
of parameters is small or similar to the dimension of the data, 
and regularization is
not needed to prevent overfitting. 
The EKI algorithm as stated preserves the linear span of the initial
ensemble $\{\theta_0^{(j)}\}_{j=1}^J$ for each $n$ and thus operates in
a finite dimensional vector space
\citep[Theorem 2.1]{iglesias2013ensemble}, 
even if $\Theta$ is an infinite dimensional
vector space; this form of regularization may be useful
for high-dimensional unknown parameters.
Other types of regularization can also be incorporated into 
EKI if needed, as detailed in the introduction; in addition
sparsity can be used to regularize as demonstrated in
\citet{schneider2020ensemble}.

\subsection{Gaussian Process Regression (GPR)}
\label{ssec:GPR}

Throughout this paper, we apply the EKI algorithm in a finite-dimensional vector space $\Theta \subseteq \bbR^p.$ 
In some cases, we use GPR to estimate unknown 
functions appearing in our SDE. 
More precisely, we will use the mean $m(x)$ of a
Gaussian process conditioned on noisy observations at a
fixed set of design points; we will then optimize 
over the observed values of the process at the
design points, over the standard deviations
of the noise in these observations
and over the parameters describing
the covariance function of the Gaussian process. This leads to a parameter dependent
function $m(x;\theta^\textrm{GP})$, the construction of which we now detail. Note that the
construction involves probabilistic considerations but, once completed, leads to
a class of deterministic functions $m(x;\theta^\textrm{GP})$ over $\theta^\textrm{GP} \in \Theta^\textrm{GP}.$
The key advantage of the GPR construction is that it leads to a hierarchical
parameterization that has proved very useful in many machine learning tasks \citep[see, e.g.,][]{bernardo1998regression,rasmussen2006gaussian}.
Adapting it to statistical estimation more generally is potentially very fruitful,
and is one of the ideas we use here. We are essentially proposing to use
Gaussian process parameterizations beyond the simple regression setting, into
more general function learning problems.

The function $m$ is defined as the minimizer of
\[L(m):=\frac12\|m\|_\mathsf{K}^2+\sum_{r=1}^R \frac{1}{2\sigma_{(r)}^2}|m(x_{(r)})-m'_r|^2\]
over $\mathsf{K}$, the reproducing kernel Hilbert space associated with covariance function
$k(x,x';a,\ell);$ here $a$ and $\ell$ represent and amplitude and lengthscale
parameter of the covariance function.
By the representer theorem \citep[see, e.g.,][]{rasmussen2006gaussian}, it follows that
the minimizer lies in the linear span of
$\{k(x,x_{(r)};a,\ell)\}_{r=1}^R\}$ and may be found by 
solving a linear system of dimension equal to $R$. The solution of this
linear system depends on the $\{m(x_{(r)}),\sigma_{(r)}\}_{r=1}^R.$

As a consequence, the set of parameters $\theta^\textrm{GP}$ defining our
parameterized function $m(x;\theta^\textrm{GP})$
contains the following elements:
\begin{enumerate}[(i)]
\item noisily observed values of $\{m(x_{(r)})\}$,  at some fixed nodes $x_{(r)}$, comprising the vector $\{m_r'\}$; 
\item observation error variances 
$\Sigma_{\textrm{obs}}=\textrm{diag}\{(\sigma_{(r)})^2\}$ 
at the fixed nodes $x_{(r)}$; 
\item hyper-parameters ($a$, $\ell$) that represents an amplitude and a length-scale of the kernel 
$k(x, x^\prime;a,\ell)$ used in the GP regression.
\end{enumerate}

In this setting $m'$ and $\Sigma_\textrm{obs}$
each contain $R$ elements. Thus $\Theta^\textrm{GP}=\mathbb{R}^{2R+2}.$ 
The result of the minimization is a complex nonlinear function
of $\theta^\textrm{GP}=(m',\Sigma_{\textrm{obs}},\sigma,\ell)
\in \Theta^\textrm{GP}.$ We use EKI to learn $\theta^\textrm{GP}$ together with other unknown parameters in the modeled systems using time-averaged statistics as data. The linear span of $\{k(x,x_{(r)};a,\ell)\}_{r=1}^R\}$ 
comprises a set of adaptive basis functions which, via the parameters 
$a$ and $\ell$, may be
adapted to the observed finite-time average data. 
The setting may be generalized or
simplified in different ways: for example $\Sigma_{\textrm{obs}}$
can be chosen to
be an arbitrary symmetric positive-definite matrix to be learned or, as we do in 
the numerical examples in this 
paper, can be chosen as $\Sigma_{\textrm{obs}}=\sigma^2 \mathrm{I}$ where $\sigma$ is
optimized over and $\mathrm{I}$ denotes the identity matrix.
The detailed settings of numerical experiments are described at the beginning of each sub-section in Section~\ref{sec:N}. Since we take 
$\Sigma_{\textrm{obs}}$ to be a constant diagonal matrix throughout this
paper, $\Theta^\textrm{GP}$ simplifies to $\mathbb{R}^{R+3}$.

\section{Numerical Results}
\label{sec:N}
To demonstrate the capability of the proposed methodology, we apply it to four different examples, including classical chaotic systems (Lorenz 63 system in subsection~\ref{ssec:NL63}, Lorenz 96 system in subsection~\ref{ssec:NL96}), climate dynamics (ENSO in subsection~\ref{ssec:NENSO}), and molecular dynamics (butane molecule dihedral angle in subsection~\ref{ssec:NButane}). All these numerical studies confirm that the proposed methodology serves well as a systematic approach to 
the fitting of SDE models to data. In particular the
data we use appears to lead to identifiable parameter estimation
problems in every example presented. In all cases,
the data are initially presented in two figures, one showing the ability of the EKI method
to fit the data, and a second showing how well the fitted SDE performs in terms
of reproducing the invariant measure of the true system. The ensemble size is chosen as 100 by default. The noise $\Gamma$ is estimated based on the ensemble of time-averaged data from the system with random initial conditions. It should be noted that the time-averaged data (which serve as training data) are obtained from a relatively short trajectory, and the invariant measures (which serve as testing data) are obtained by simulating the modeled system for a much longer time. We then show other figures that differ from case to case and are designed to illustrate the quality and 
nature of the fitted SDE model.

\subsection{Lorenz 63 System}
\label{ssec:NL63}

\subsubsection{Noisy Lorenz 63: Simulation Study}
\label{sssec:sim}
The first set of experiments are simulation studies in which data from \eqref{eq:l63n} is
used to fit parameters within the following model:
\begin{equation}
\begin{aligned}
\frac{dx_1}{dt}&=\alpha(x_2-x_1)+\sqrt{\sigma}\frac{dW_1}{dt} \\
\frac{dx_2}{dt}&=x_1 (\rho-x_3)-g_L(x_2)+\sqrt{\sigma}\frac{dW_2}{dt} \\
\frac{dx_3}{dt}&=x_1x_2-\beta x_3+\sqrt{\sigma}\frac{dW_3}{dt}.
\end{aligned}
\end{equation}

\begin{figure}[!htbp]
  \centering
  \includegraphics[width=0.2\textwidth]{G_legend}\\
  \subfloat[ODE]{\includegraphics[width=0.49\textwidth]{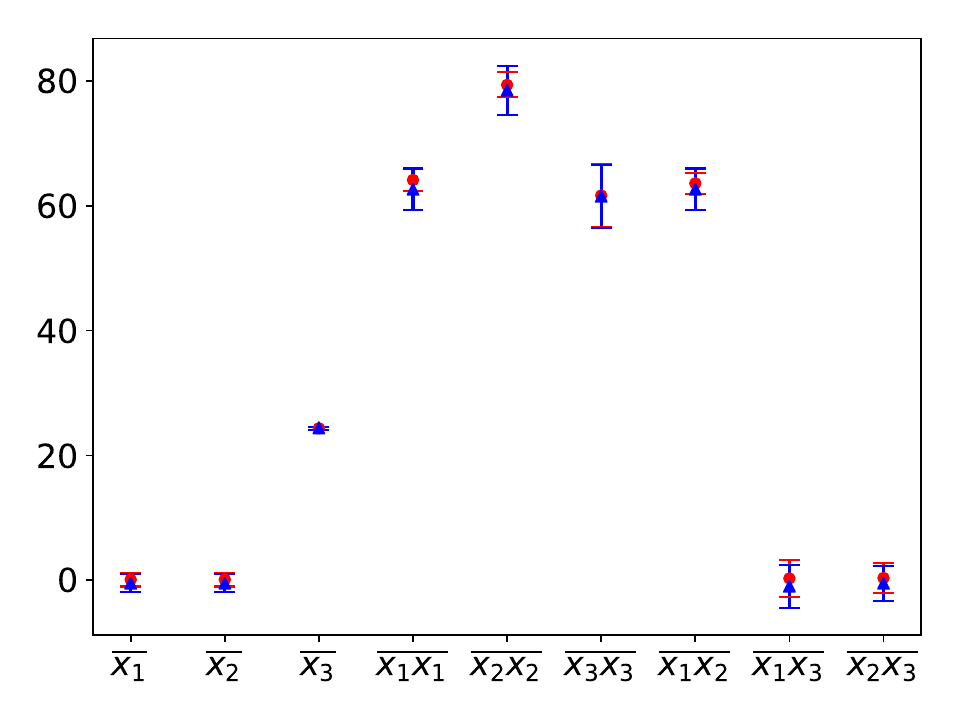}}
  \subfloat[SDE]{\includegraphics[width=0.49\textwidth]{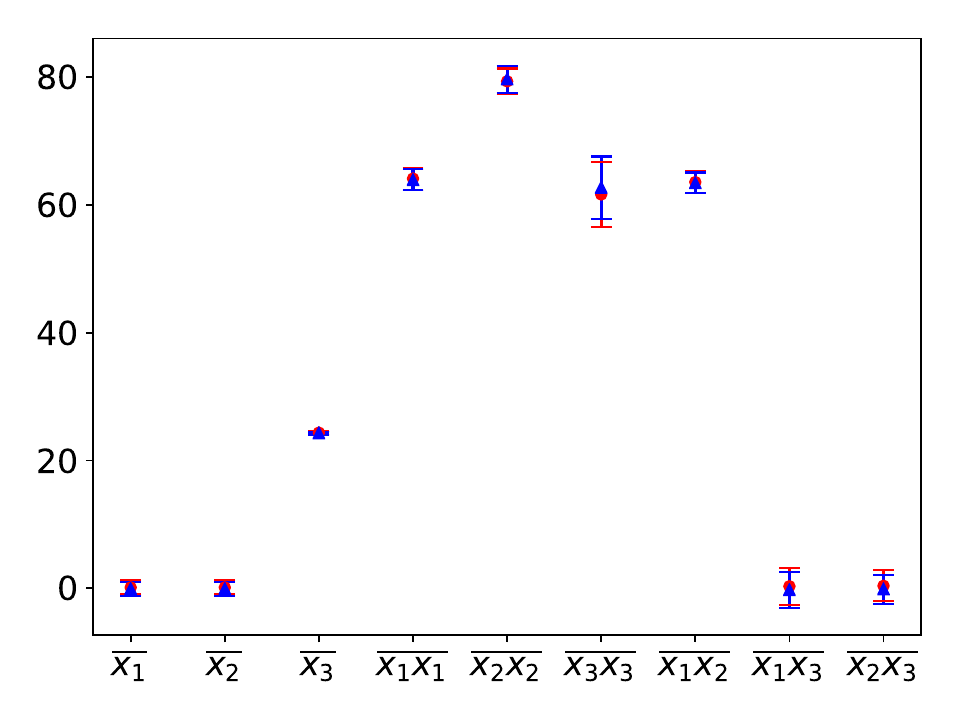}}
    \caption{First two moments of state $x$ for the Lorenz 63 system by using EKI to estimate $\sigma$ and the linear function $g_L(x_2)$.}
  \label{fig:G-L63-linear-error}
\end{figure}

\begin{figure}[!htbp]
  \centering
  \includegraphics[width=0.2\textwidth]{measure_legend}\\
  \subfloat[$x_1$ (ODE)]{\includegraphics[width=0.32\textwidth]{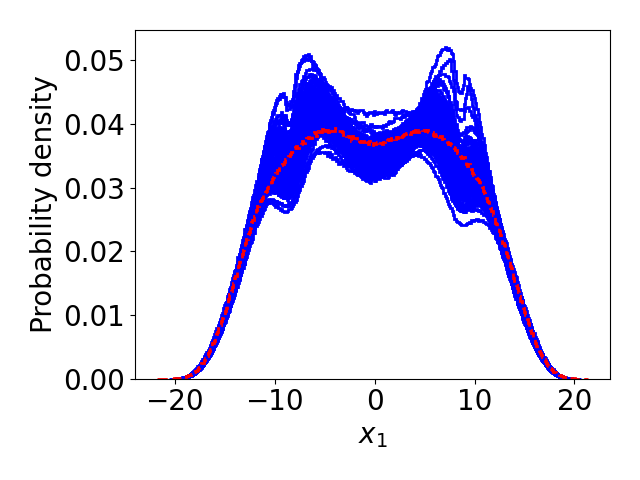}}
  \subfloat[$x_2$ (ODE)]{\includegraphics[width=0.32\textwidth]{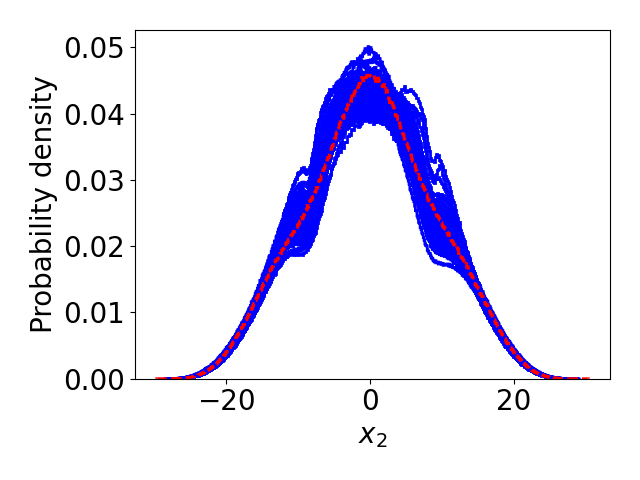}}
  \subfloat[$x_3$ (ODE)]{\includegraphics[width=0.32\textwidth]{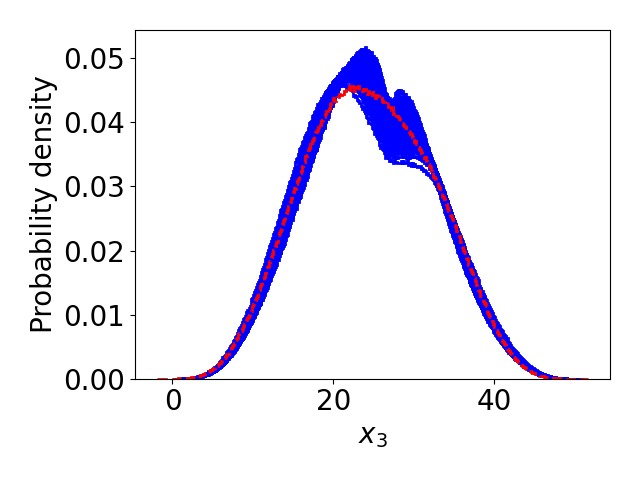}} \\
  \subfloat[$x_1$ (SDE)]{\includegraphics[width=0.32\textwidth]{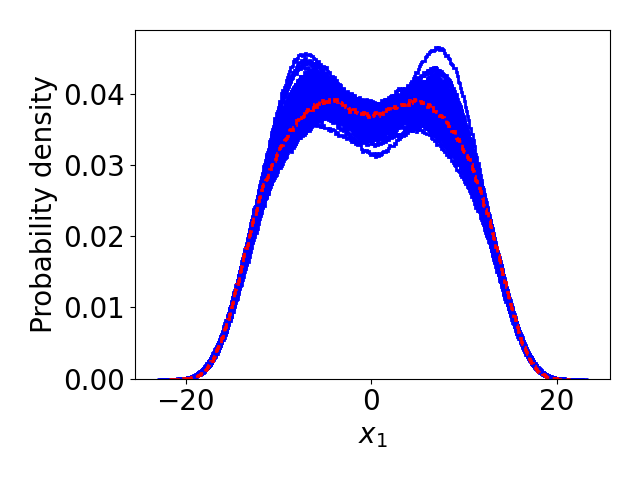}}
  \subfloat[$x_2$ (SDE)]{\includegraphics[width=0.32\textwidth]{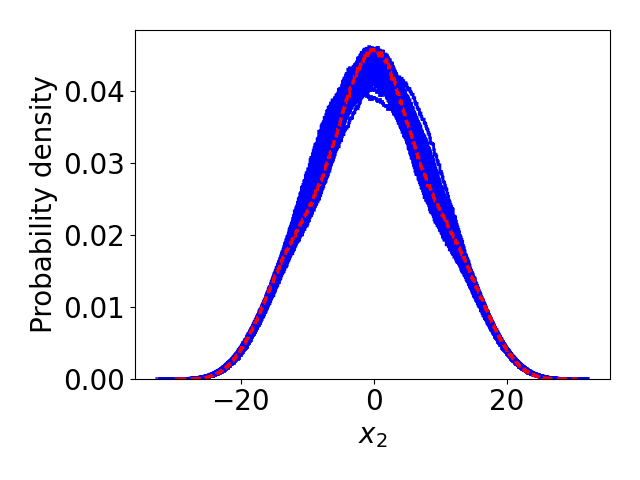}}
  \subfloat[$x_3$ (SDE)]{\includegraphics[width=0.32\textwidth]{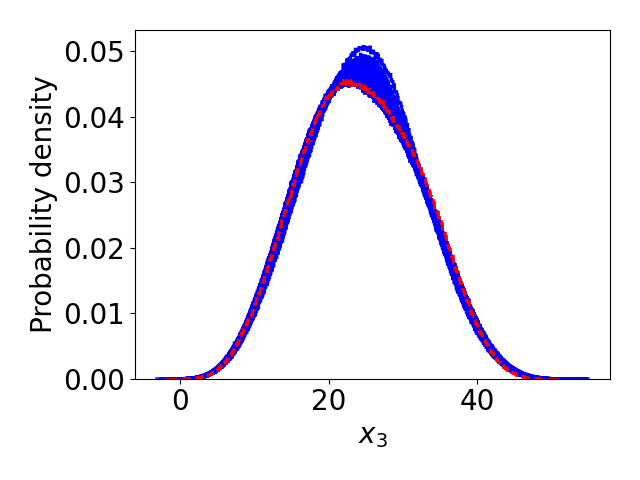}}
    \caption{Invariant measures of fitted models (with linear function $g_L(x_2)$ and $\sigma$ as unknowns) and the true noisy Lorenz 63 system.}
  \label{fig:Invariant-measure-L63-sigma_10-linear-error}
\end{figure}

\begin{figure}[!htbp]
  \centering
  \includegraphics[width=0.32\textwidth]{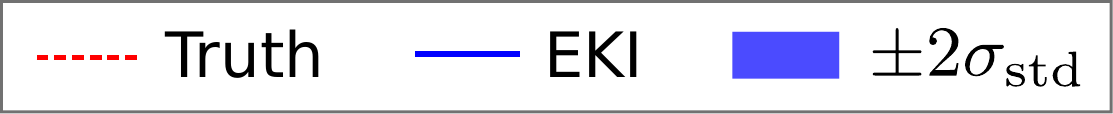}\\
  \subfloat[ODE]{\includegraphics[width=0.47\textwidth]{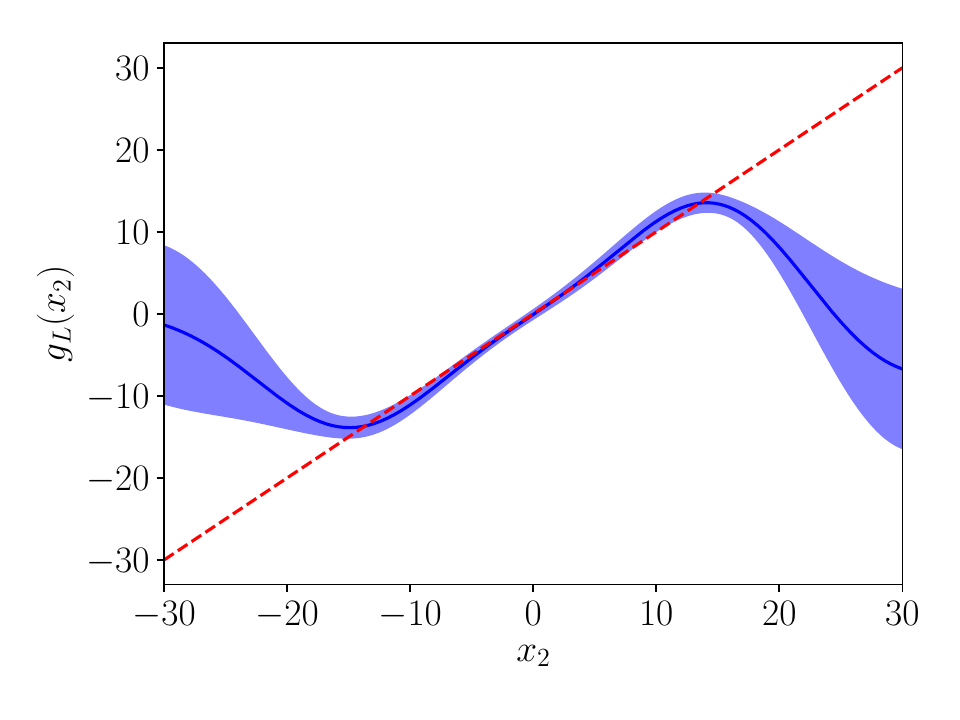}}
  \subfloat[SDE]{\includegraphics[width=0.47\textwidth]{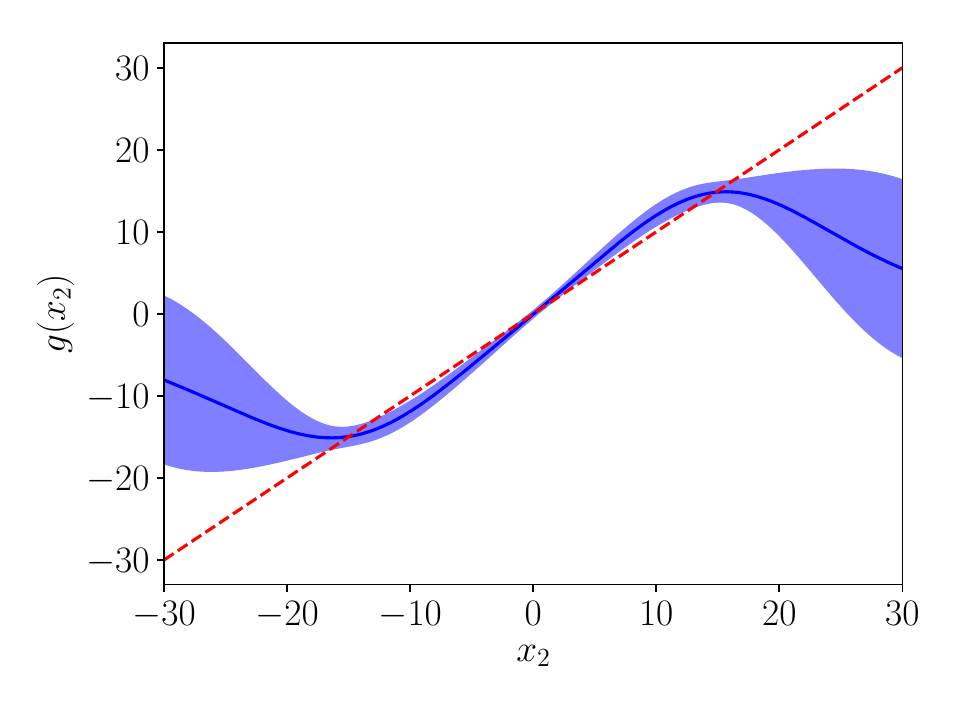}}
    \caption{The linear function $g_L(x_2)$ learnt in (a) ODE model and (b) SDE model.}
  \label{fig:L63_GP}
\end{figure}

Unlike the other examples in this paper, this initial simulation study involves data that come directly from an SDE within the model class being fitted, 
enabling a clear verification of the proposed methodology,
and confirming that the time-averaged data lead to an identifiable model.
Specifically, the data are obtained by simulating the model with a given set of parameters: $\alpha=10$, $\rho=28$, $\beta=8/3$, and $\sigma=10$, as well
as the choice $g_L(x_2)=x_2.$ Using EKI, we will fit $\theta$ defined in two different ways:
\begin{enumerate}[(i)]
\item Fix $g_L(x_2)=x_2$ and learn $\theta=(\alpha,\sigma).$ The initial ensemble of $\alpha$ and $\sqrt{\sigma}$ are uniformly drawn from $[1,20]$ and $[0.1,15]$..
\item Parameterize $g_L(x)$ by $\theta_1$, as a GP, and learn $\theta=(\theta_1,\sigma)$. The initial ensemble of $\sqrt{\sigma}$ is uniformly drawn from $[0.1,15]$. The fixed GP nodes are five points uniformly distributed in $[-30,30]$. The initial ensemble of noisily observed values on those nodes are uniformly drawn from $[-20,20]$. The initial ensemble of GP observation error is uniformly drawn from $[0.1,10]$, and the initial ensemble of GP hyper-parameters $a$ and $\ell$ are uniformly drawn from $[0.1,10]$ and $[5,10]$. Results are presented in Figs.~\ref{fig:G-L63-linear-error} to~\ref{fig:L63_GP}.
\end{enumerate}

\begin{figure}[!htbp]
  \centering
  \includegraphics[width=0.2\textwidth]{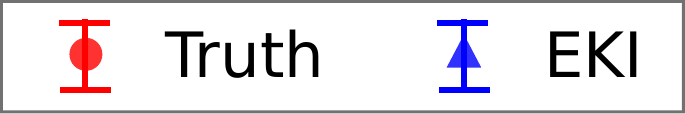}\\
  \subfloat[ODE]{\includegraphics[width=0.49\textwidth]{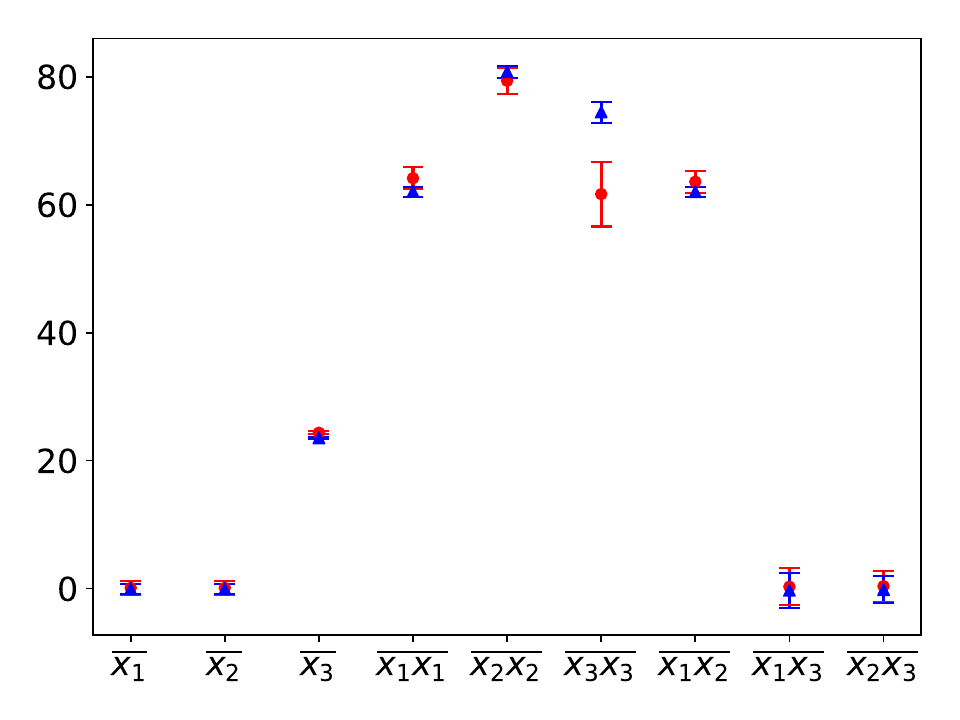}}
  \subfloat[SDE]{\includegraphics[width=0.49\textwidth]{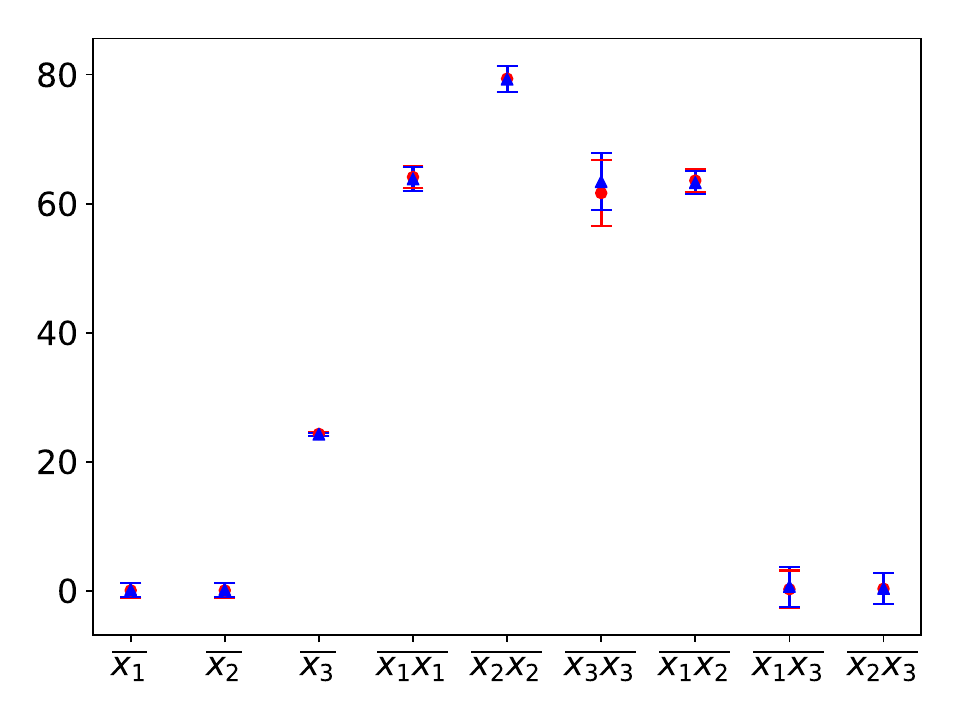}}
    \caption{First two moments of state $x$ for the Lorenz 63 system found by using EKI to estimate
    $\alpha$ (ODE case) and $\sigma,\alpha$ (SDE case).}
  \label{fig:G-L63-alpha-sigma_10}
\end{figure}

\begin{figure}[!htbp]
  \centering
  \includegraphics[width=0.2\textwidth]{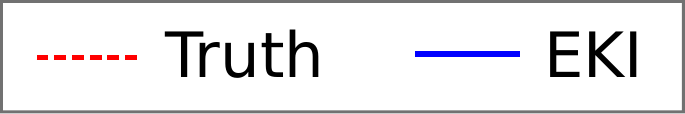}\\
  \subfloat[$x_1$ (ODE)]{\includegraphics[width=0.33\textwidth]{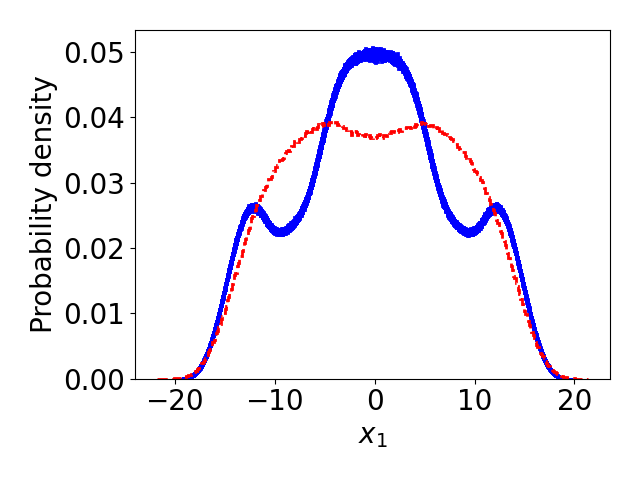}}
  \subfloat[$x_2$ (ODE)]{\includegraphics[width=0.33\textwidth]{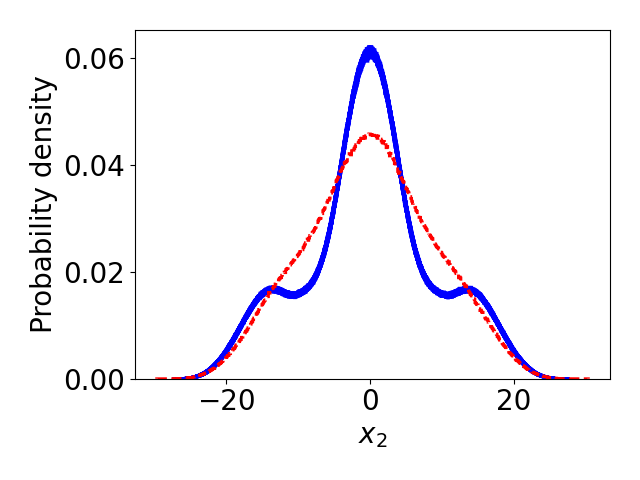}}
  \subfloat[$x_3$ (ODE)]{\includegraphics[width=0.33\textwidth]{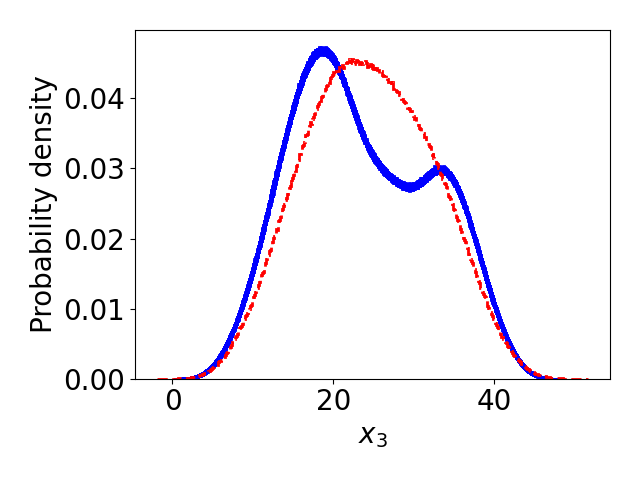}} \\
  \subfloat[$x_1$ (SDE)]{\includegraphics[width=0.33\textwidth]{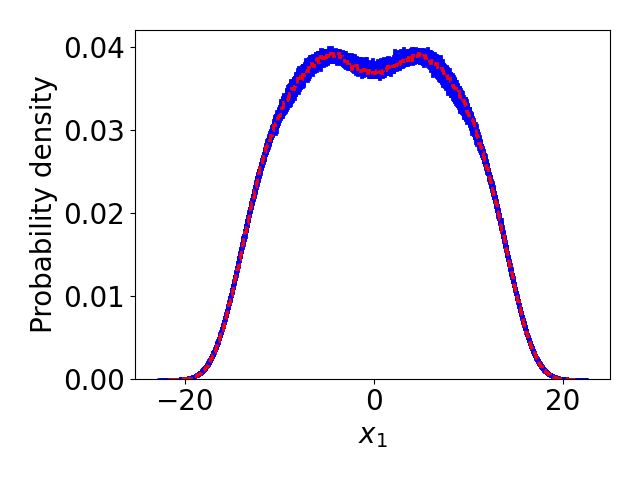}}
  \subfloat[$x_2$ (SDE)]{\includegraphics[width=0.33\textwidth]{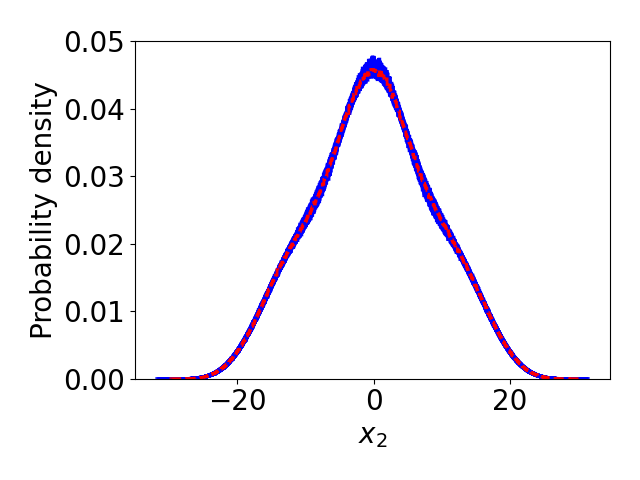}}
  \subfloat[$x_3$ (SDE)]{\includegraphics[width=0.33\textwidth]{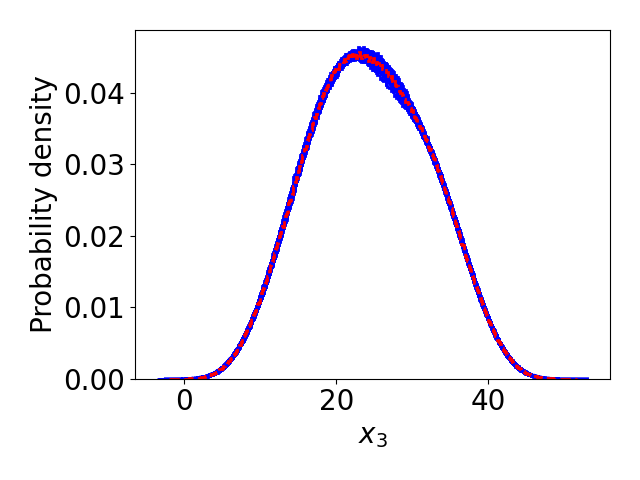}}
    \caption{Invariant measures of fitted models compared with those of the true noisy Lorenz 63 system: 
    ODE case, fit $\alpha$; SDE case, fit $\sigma,\alpha$.}
  \label{fig:Invariant-measure-L63-sigma_10}
\end{figure}

The trajectory initial condition
is uniformly drawn from $[0,1)$ 
for each state variable. 20 EKI iterations is used. Results are presented in Figs.~\ref{fig:G-L63-alpha-sigma_10} to~\ref{fig:Invariant-measure-L63-sigma_10}. The data in this case are a finite-time average approximation of $\{\cG_1(x),\cG_2(x)\}$, i.e., the first and second moments of the state vector $x$ are used as observational data. Therefore, the data vector $y$ has 9 elements in total. 
In case (i), we fit both an ODE (constrain $\sigma=0$ and learn $\alpha$) 
and an SDE (learn both $\alpha$ and $\sigma$)
to the given data.  
We are fitting $1$ parameter in the ODE case and $2$ parameters 
in the SDE setting, in both cases using a data vector $y$ of dimension $9$. The comparison of the
output of the EKI algorithm with the true data, in the case of both ODE and SDE fits,
is presented in Fig.~\ref{fig:G-L63-alpha-sigma_10}.
In the ODE case, the algorithm fails to match one second moment (left panel), while
in the SDE case all moments are well-matched. This has a significant effect
on the ability to reproduce the invariant measure (Fig.~\ref{fig:Invariant-measure-L63-sigma_10}). In the upper row (ODE), the fit is very
poor whereas in the lower row (SDE) it is excellent. 
The fit of the SDE is not surprising since the data is generated directly 
from within the model class to be fit; the behavior of the fit to an ODE
gives insight into the method when the model is misspecified.

\begin{figure}[!htbp]
  \centering
  \includegraphics[width=0.2\textwidth]{measure_legend}\\
  \subfloat[$x_1$ (ODE)]{\includegraphics[width=0.44\textwidth]{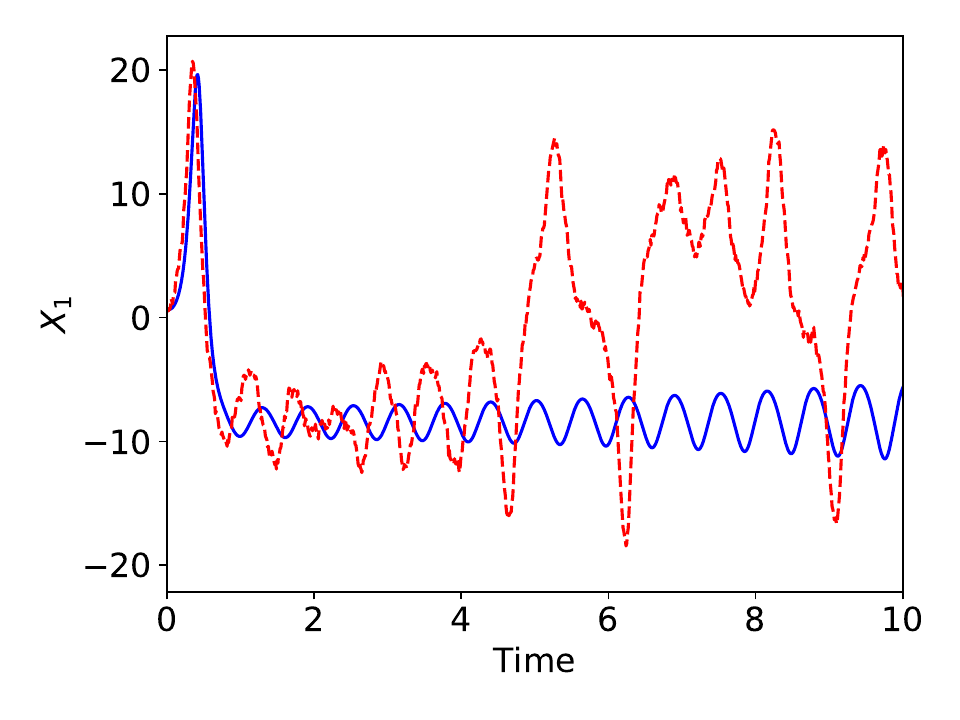}}
  \subfloat[$x_1$ (SDE)]{\includegraphics[width=0.44\textwidth]{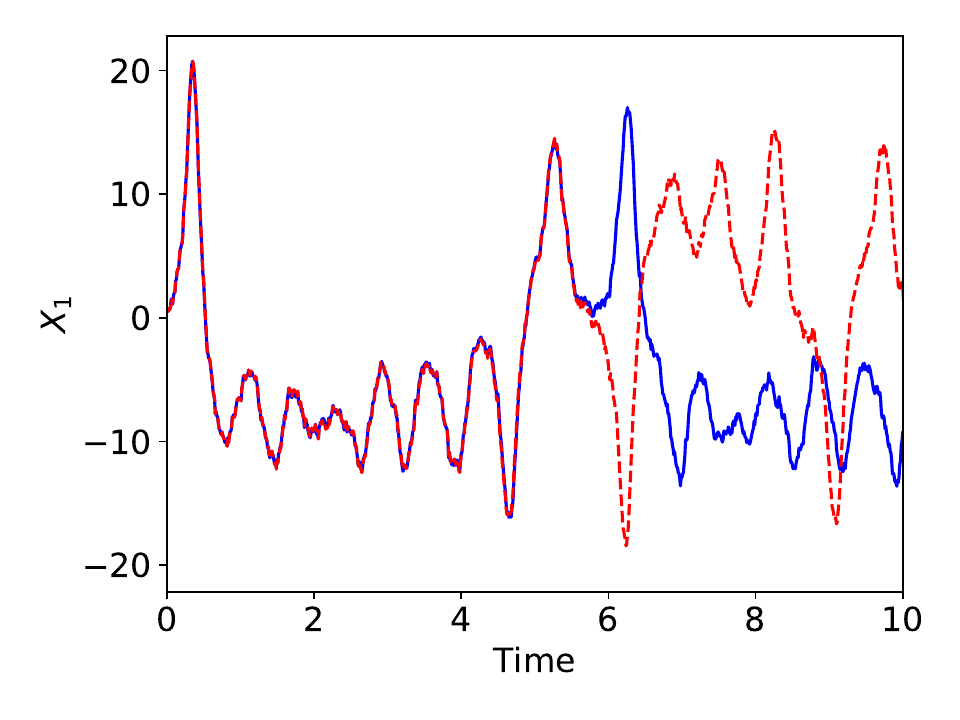}} \\
  \subfloat[RMSE (ODE)]{\includegraphics[width=0.44\textwidth]{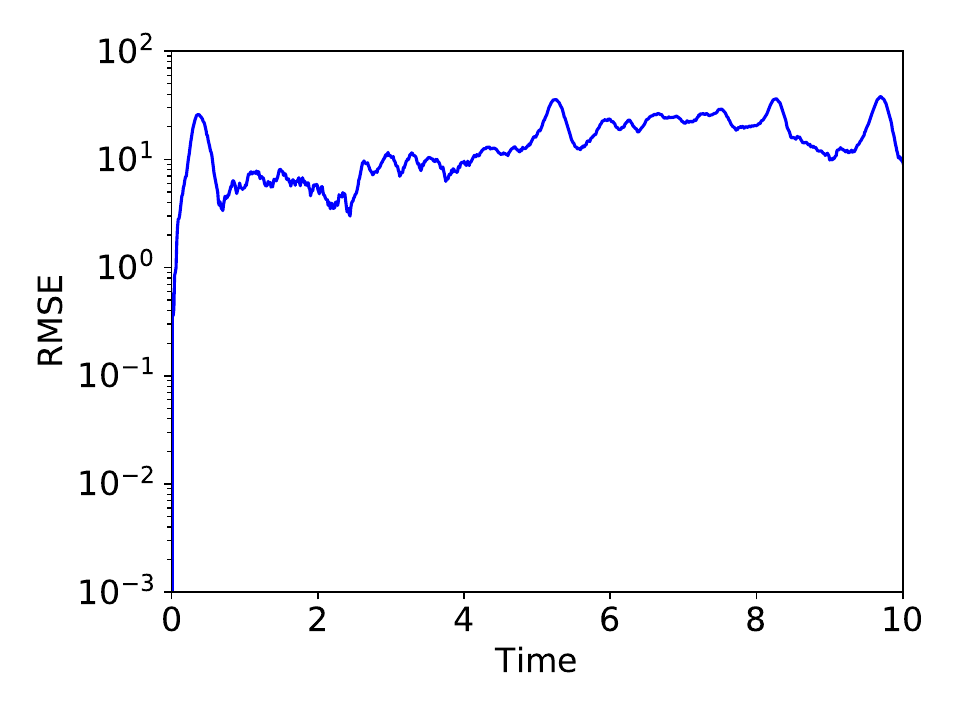}} 
  \subfloat[RMSE (SDE)]{\includegraphics[width=0.44\textwidth]{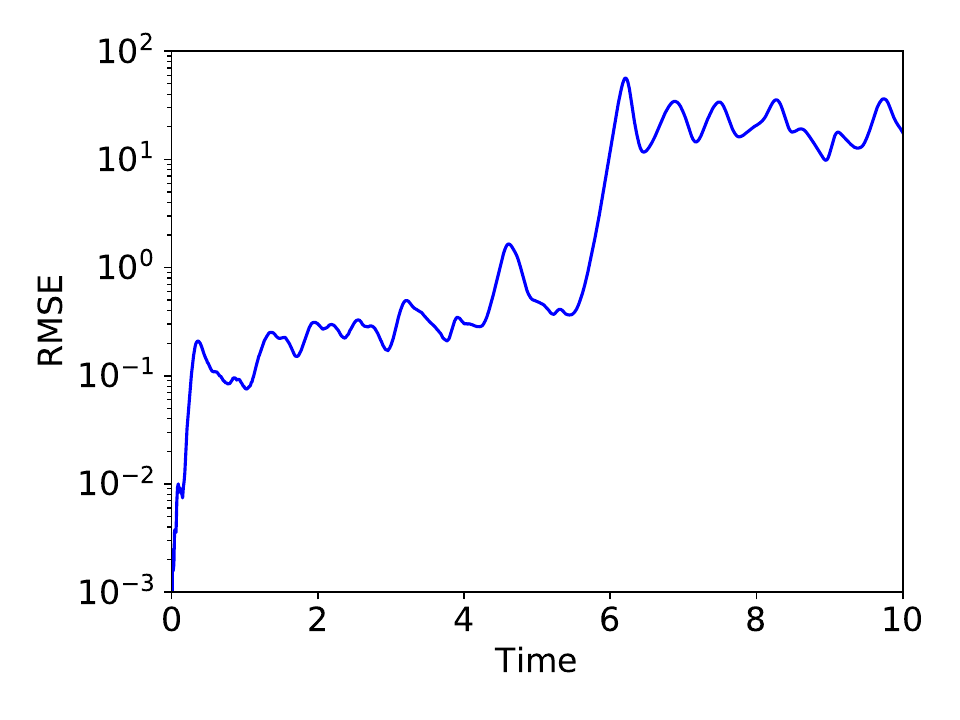}}
    \caption{State trajectory and RMSE of fitted models compared with those of the true noisy Lorenz 63 system: 
    ODE case, fit $\alpha$; SDE case, fit $\sigma,\alpha$. The results are obtained by fitted models with ensemble mean of estimated parameters.}
  \label{fig:Trajectory-L63-sigma_10}
\end{figure}

Figure~\ref{fig:Trajectory-L63-sigma_10} presents the trajectory of $x_1$ and 
the RMSE calculated using all state variables. To facilitate the comparison 
against the true system, we use the same initial condition, and the same 
random seed for the stochastic process, for both true and modeled systems. 
It should be noted that the training of the modeled systems does not make 
use of trajectory information from the true system. As shown in 
Fig.~\ref{fig:Trajectory-L63-sigma_10}, the modeled system with the stochastic 
term demonstrates better performance in matching the trajectory of the true 
system.

\begin{figure}[!htbp]
  \centering
  \includegraphics[width=0.2\textwidth]{measure_legend}\\
  \subfloat[ODE]{\includegraphics[width=0.44\textwidth]{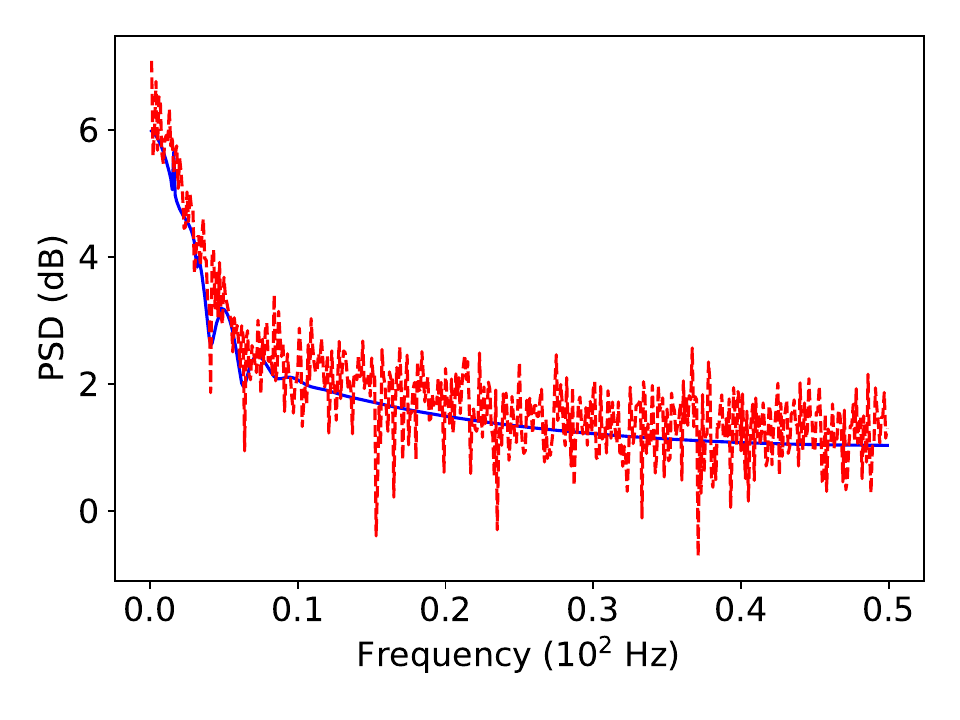}}
  \subfloat[SDE]{\includegraphics[width=0.44\textwidth]{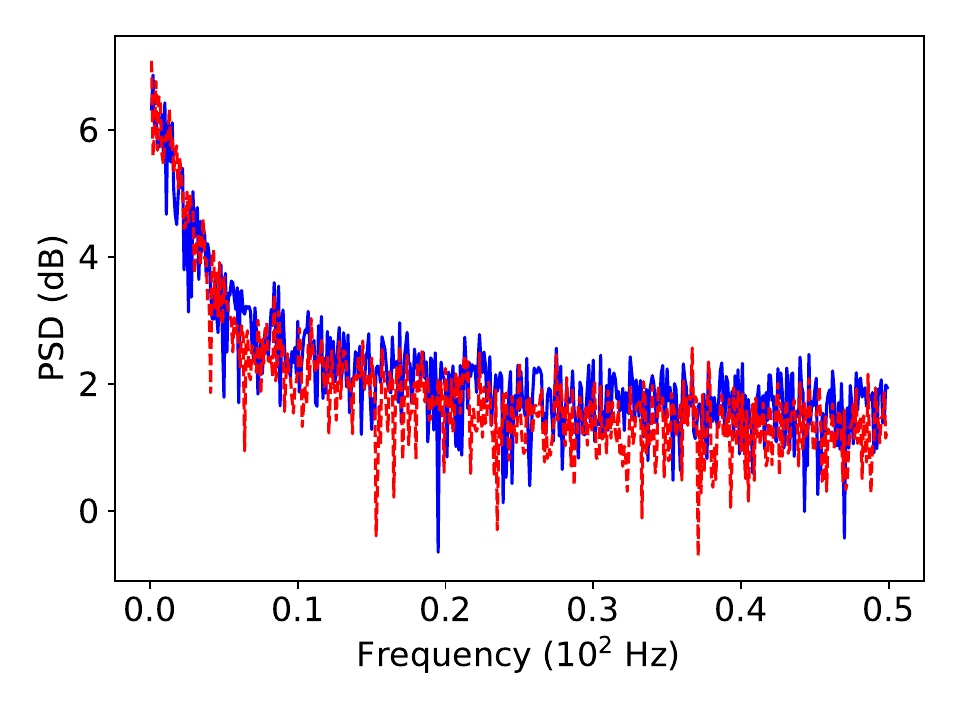}}
    \caption{Power spectral density of fitted models compared with those of the true noisy Lorenz 63 system: 
    ODE case, fit $\alpha$; SDE case, fit $\sigma,\alpha$. The results are obtained by fitted models with ensemble mean of estimated parameters.}
  \label{fig:PSD-L63-sigma_10}
\end{figure}

The comparison of the power spectral density (PSD) is presented in Fig.~\ref{fig:PSD-L63-sigma_10}. Although the PSD is not used as data in this example, both fitted models capture the general pattern of the PSD of the true system. For more complex systems, PSD can provide additional information to help better identify the modeled system, and we demonstrate the use of PSD as part of data in Sections \ref{ssec:NENSO} and \ref{ssec:NButane}.

We now perform exactly the same set of experiments, fitting both an ODE and an SDE,
but allowing the function $g_L$ to be learnt as well. The function $g_L$ is parameterized 
by the mean of a GP (with unknown values specified at $5$ fixed nodes, and 
unknown constant observation error and two unknown hyper-parameters). Thus we are fitting $8$ parameters for the ODE and $9$ parameters for the SDE,
using a data vector $y$ of dimension $9$. When we do this, we are able
to obtain a good the fit for the ODE in both data space, seen in 
Fig.~\ref{fig:G-L63-linear-error}a, and as measured by the fit to the
invariant measure, as see in Fig.~\ref{fig:Invariant-measure-L63-sigma_10-linear-error}a-c.
Nonetheless the fit achieved by using an SDE remains substantially better,
as seen in Fig.~\ref{fig:G-L63-linear-error}b and Fig.~\ref{fig:Invariant-measure-L63-sigma_10-linear-error}d-f. The ensemble means of the Gaussian process learnt in the ODE and SDE models are presented in Fig.~\ref{fig:L63_GP}. Both GPs successfully capture the linear trend in the middle range of $x_2$, while the GP learnt in the ODE model exhibits
more oscillations around the middle part and more rapid deviation from the linear trend at both ends.

\subsubsection{Deterministic Lorenz 63: Dimension Reduction}
\begin{figure}[!htbp]
  \centering
  \includegraphics[width=0.2\textwidth]{G_legend}\\
  \subfloat[Moments]{\includegraphics[width=0.75\textwidth]{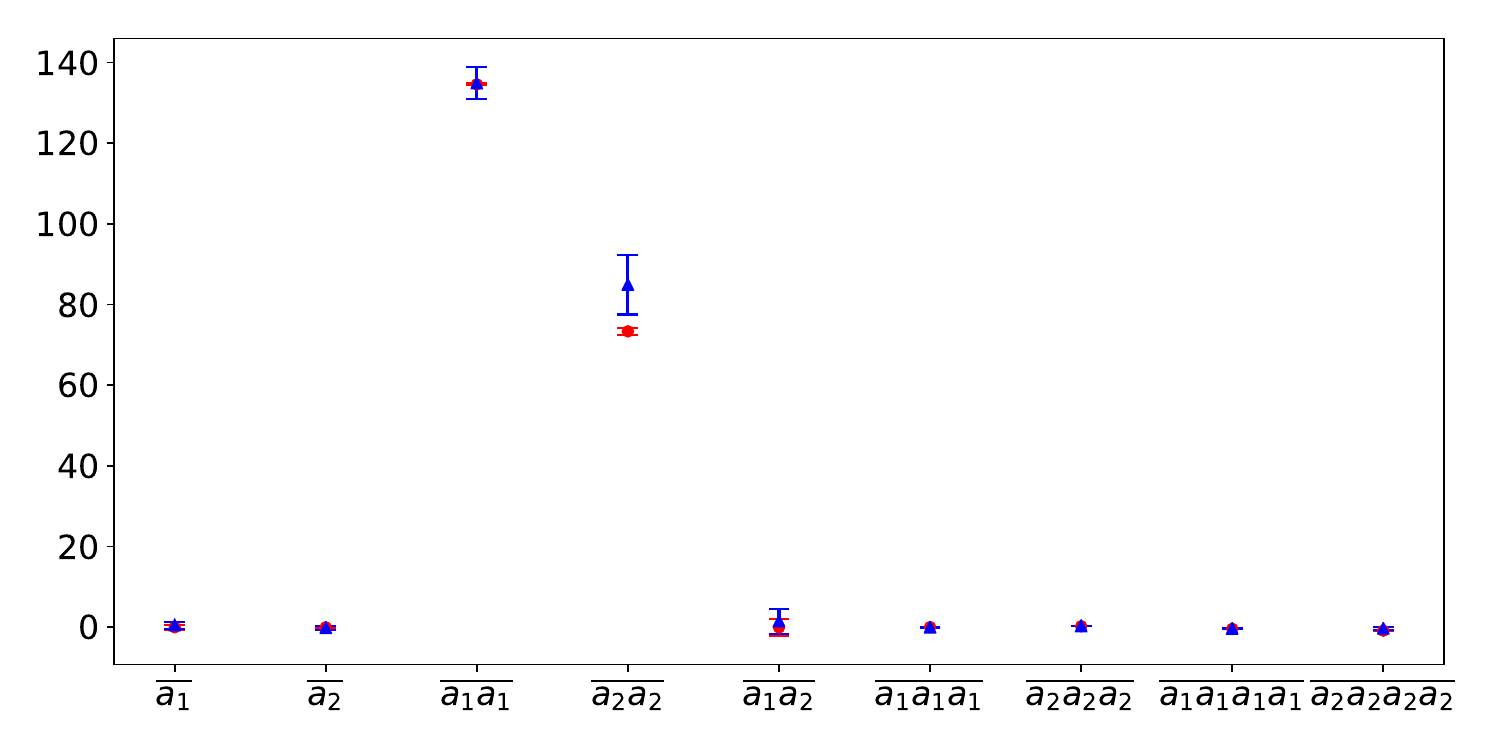}} \\
  \subfloat[Autocorrelation ($a_1$)]{\includegraphics[width=0.49\textwidth]{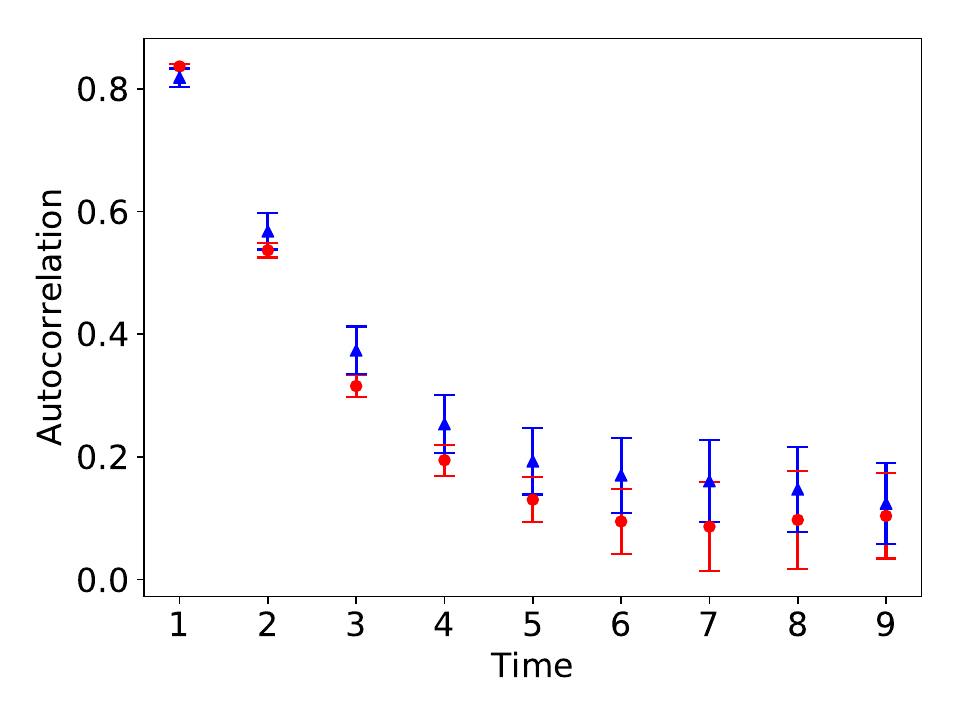}}
  \subfloat[Autocorrelation ($a_2$)]{\includegraphics[width=0.49\textwidth]{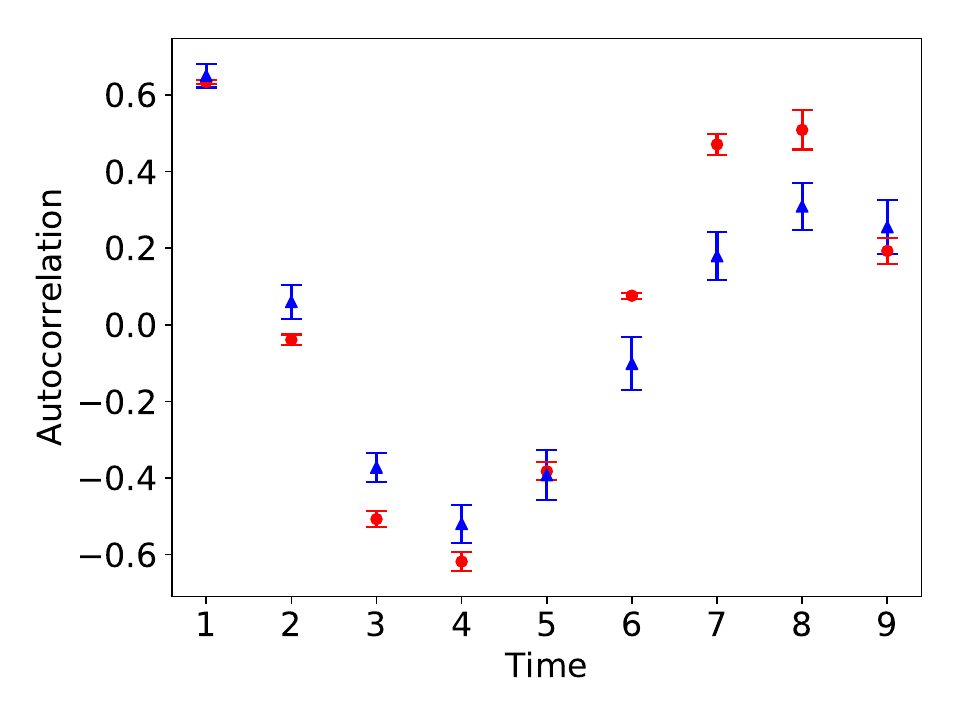}}  
    \caption{Comparison of observation data between the true system and the EKI-fitted SDE model for the deterministic Lorenz 63 system.}
  \label{fig:G-deter-L63}
\end{figure}

\begin{figure}[!htbp]
  \centering
  \includegraphics[width=0.2\textwidth]{measure_legend}\\
  \subfloat[$a_1$]{\includegraphics[width=0.44\textwidth]{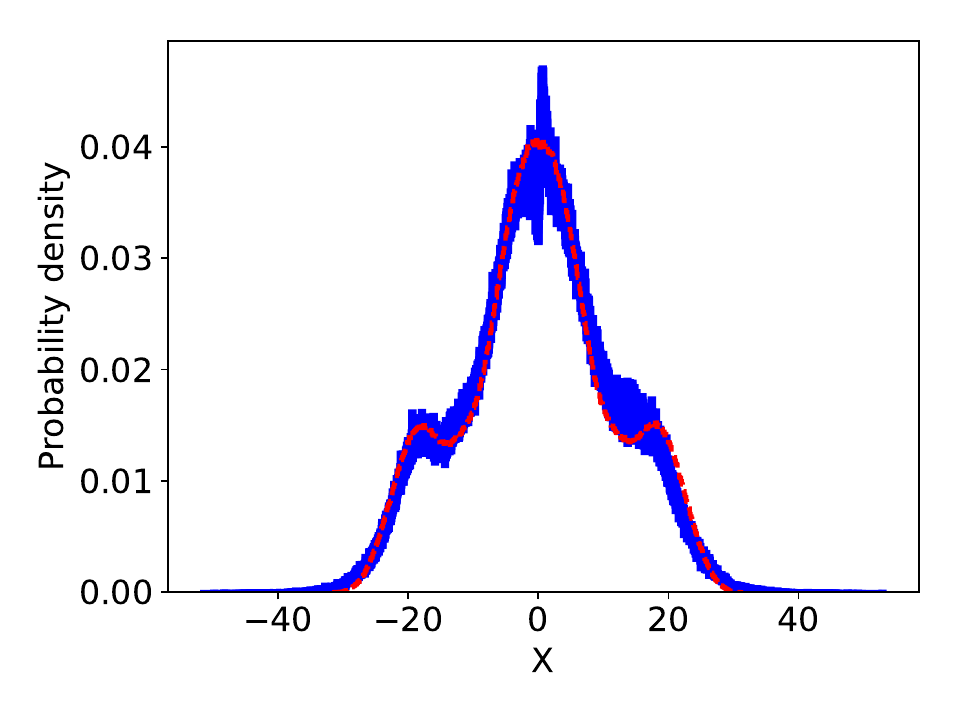}}
  \subfloat[$a_2$]{\includegraphics[width=0.44\textwidth]{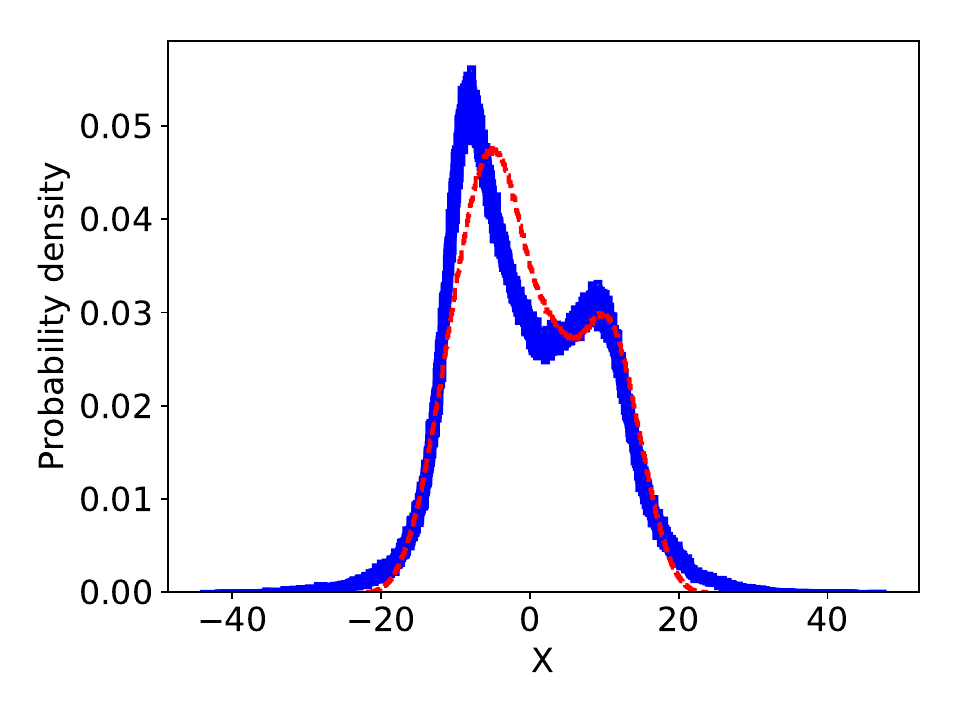}}
    \caption{Invariant measures of the deterministic Lorenz 63 system and the fitted reduced-order SDE model.}
  \label{fig:Invariant-measure-deter-L63}
\end{figure}

\begin{figure}[!htbp]
  \centering
  \subfloat[$a_1$ (truth)]{\includegraphics[width=0.49\textwidth]{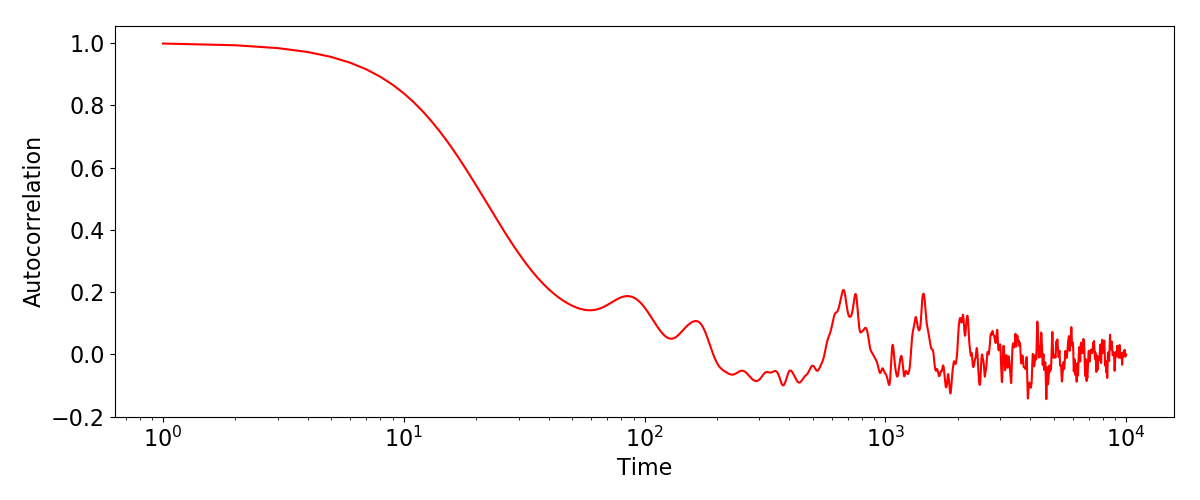}}
  \subfloat[$a_2$ (truth)]{\includegraphics[width=0.49\textwidth]{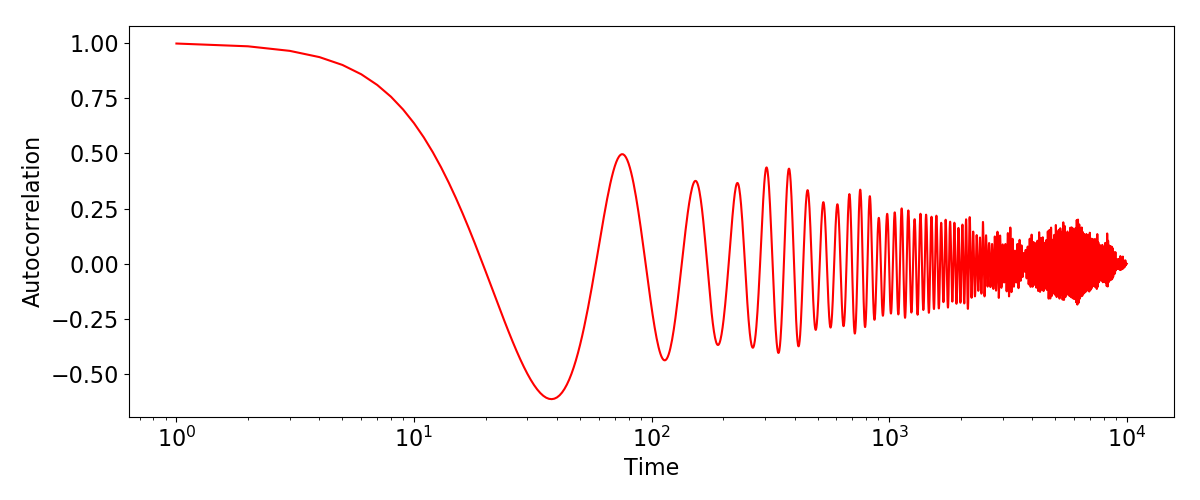}}\\
  \subfloat[$a_1$ (EKI)]{\includegraphics[width=0.49\textwidth]{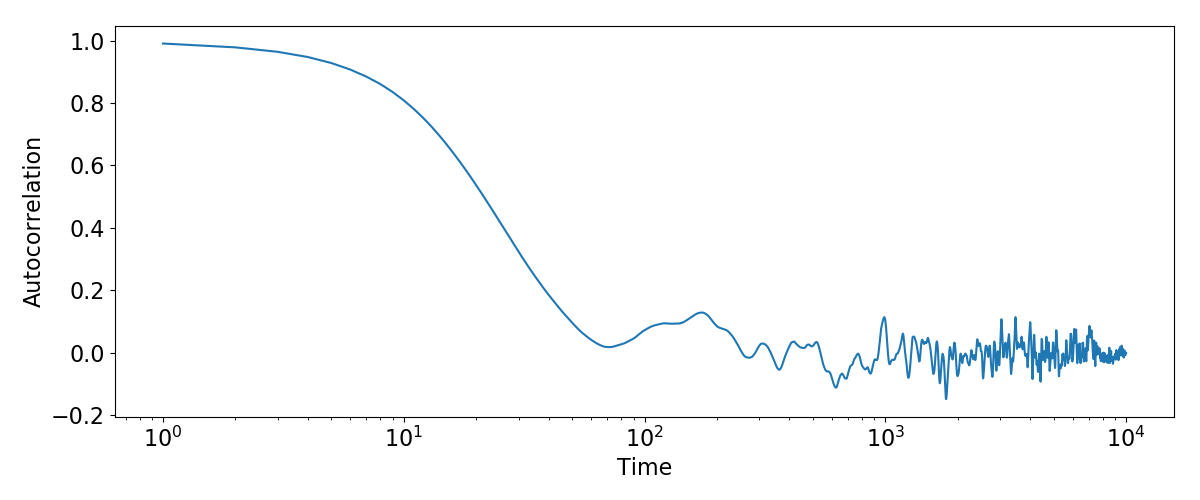}}
  \subfloat[$a_2$ (EKI)]{\includegraphics[width=0.49\textwidth]{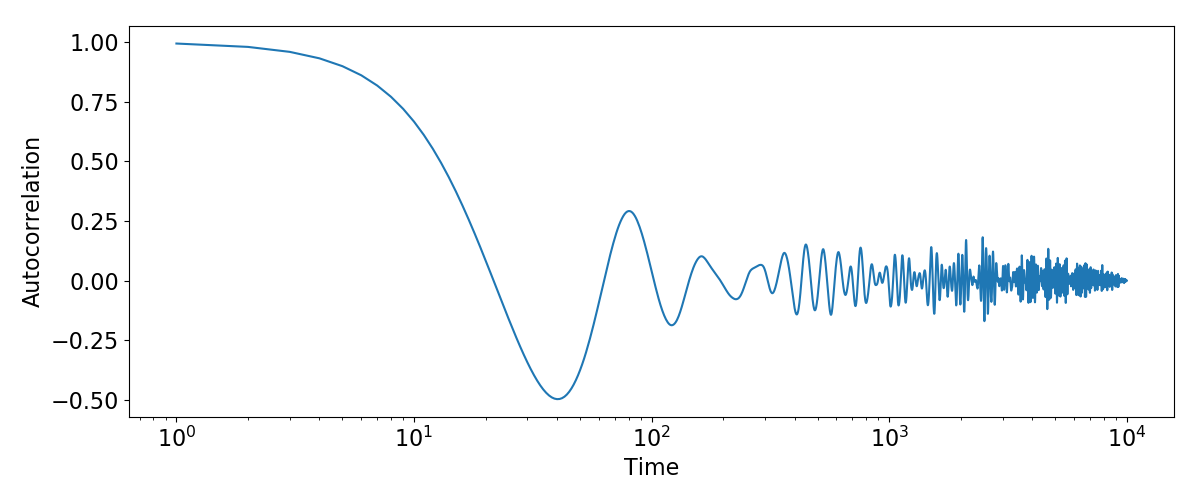}}
    \caption{Autocorrelation of the deterministic Lorenz 63 system and the fitted reduced-order SDE model.}
  \label{fig:autocorr-deter-L63}
\end{figure}

\begin{figure}[!htbp]
  \centering
  \subfloat[$a_1$ (truth)]{\includegraphics[width=0.49\textwidth]{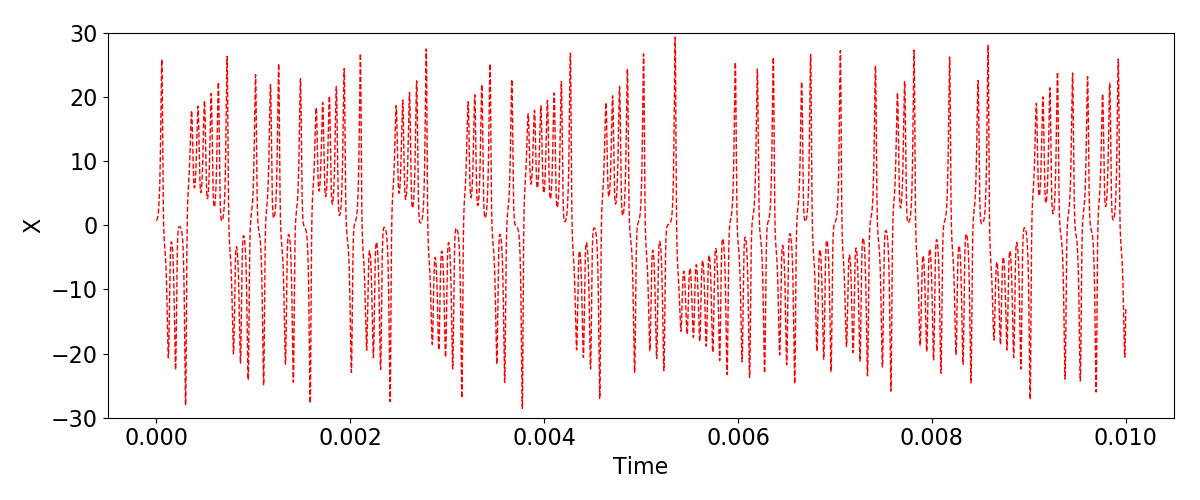}}
  \subfloat[$a_2$ (truth)]{\includegraphics[width=0.49\textwidth]{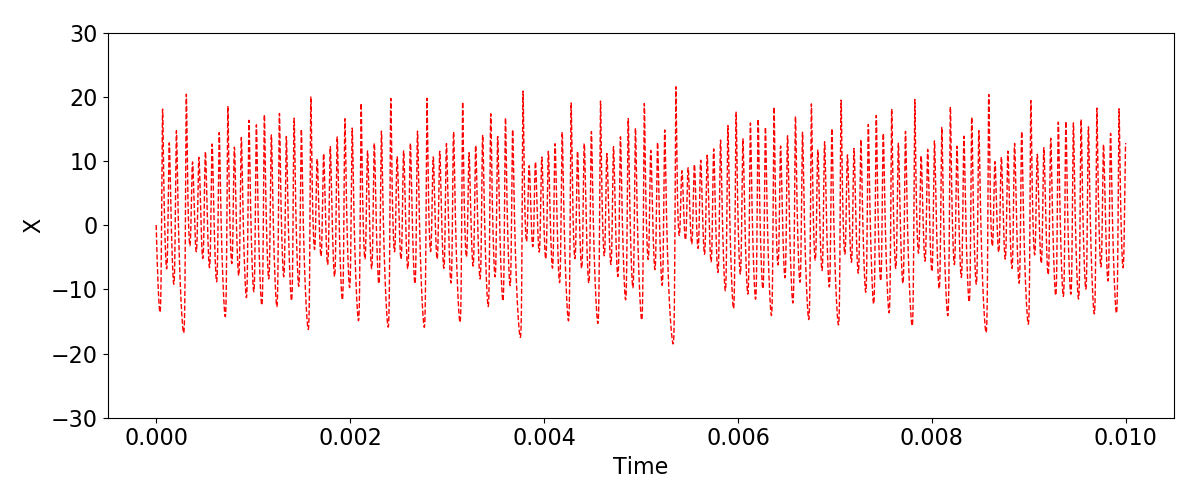}}\\
  \subfloat[$a_1$ (EKI)]{\includegraphics[width=0.49\textwidth]{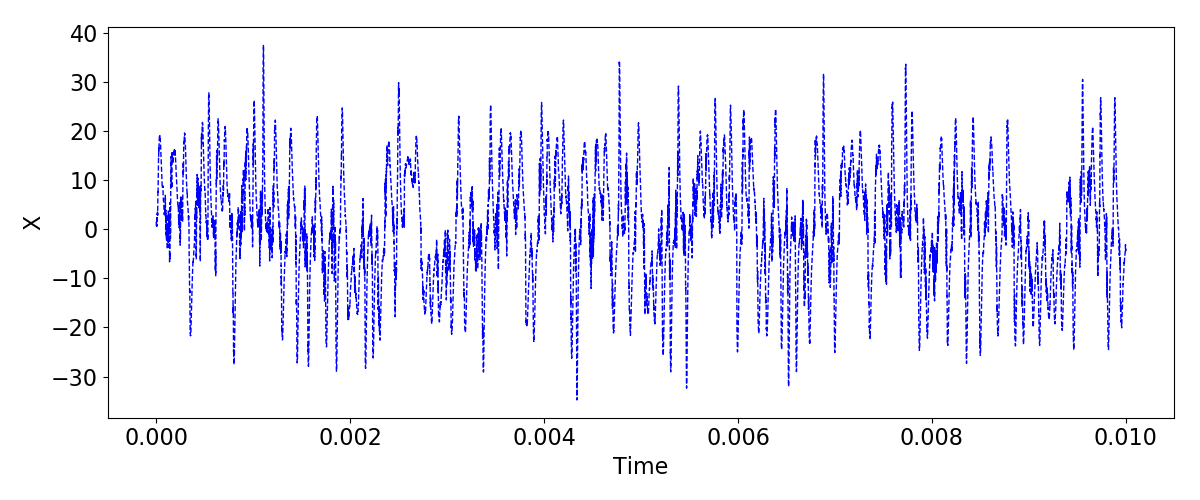}}
  \subfloat[$a_2$ (EKI)]{\includegraphics[width=0.49\textwidth]{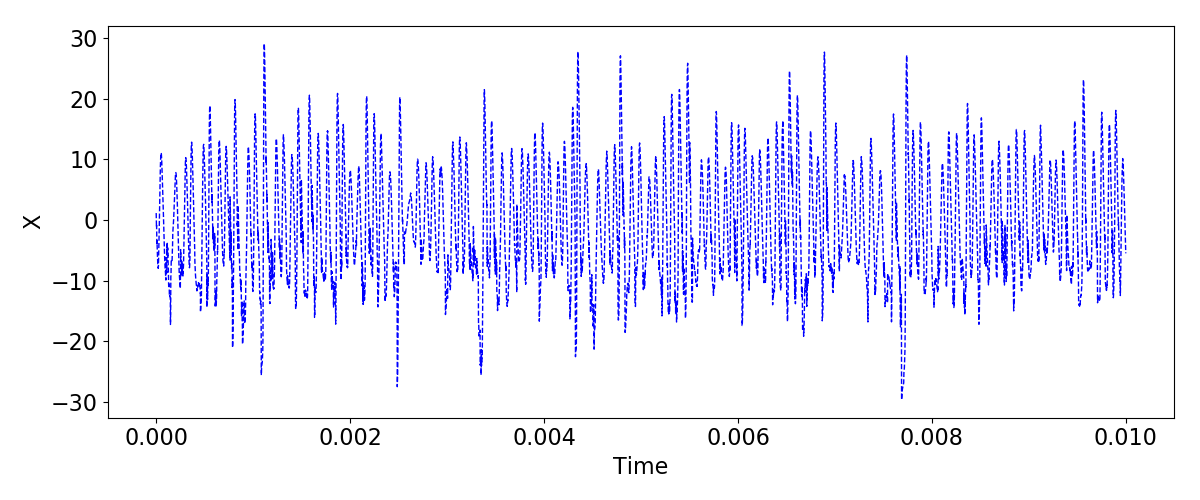}}
    \caption{Time series of the deterministic Lorenz 63 system and the fitted reduced-order SDE model.}
  \label{fig:time-series-deter-L63}
\end{figure}

 We fit a two-dimensional SDE model of the form \eqref{eq:l63pcat} to data generated
from the first two components of the three-dimensional ODE model \eqref{eq:l63pca}, \eqref{eq:gl63}. The four unknown functions ($\psi_1(\cdot)$, $\psi_2(\cdot)$, $\sigma_1(\cdot)$, $\sigma_2(\cdot)$) are parameterized by the mean of GPR. Each GPR involves the mean values at five fixed nodes, a stationary observation error, and the stationary hyper-parameters $(\sigma_{\mathcal{GP}}, \ell)$. The precise form of the data in this case is a finite time
average approximation of $\{\cG_m(a),\cG_{ac}(a)\}$. Specifically, the first four moments of the state vector $a$ are used in this case. In addition, nine equally spaced
data points are used from a finite time
average approximation of the autocorrelation function $\cG_{ac}(a)$ for both states $a_1$ and $a_2$,
from an interval of $10$ time units, and we obtain $18$ data points in total from the autocorrelation function. Thus we are fitting $32$ parameters for the SDE by using a data vector $y$ of dimension $27$. The fixed GP nodes are five points uniformly distributed in $[-30,30]$. The initial ensemble of noisily observed values on those nodes are uniformly drawn from $[-20,20]$. The initial ensemble of GP observation error is uniformly drawn from $[0.1,10]$, and the initial ensemble of GP hyper-parameters $a$ and $\ell$ are uniformly drawn from $[0.1,10]$ and $[5,10]$. The trajectory initial condition is uniformly drawn from $[0,1)$ for each state variable. We use 30 EKI iterations.
The true moment data and autocorrelation data are presented in Fig.~\ref{fig:G-deter-L63}, alongside
the results obtained by using EKI to fit the SDE model to this data, showing a relatively good match 
to the data. When measured by the ability to reproduce the invariant measure, the results
demonstrate a strong match between the marginal invariant densities on $a_1$
and $a_2$ from the original three-dimensional ODE and from the fitted two-dimensional
SDE (Fig.\ref{fig:Invariant-measure-deter-L63}). 
The fit to the autocorrelation functions of $a_1$ and $a_2$ is also quite good (Fig.~\ref{fig:autocorr-deter-L63}). It should be noted that Fig.~\ref{fig:G-deter-L63} presents the autocorrelation at discretized time shifts where we have training data, and Fig.~\ref{fig:autocorr-deter-L63} presents the autocorrelation which also extends beyond the time shifts used in training. Figure~\ref{fig:time-series-deter-L63} compares
the trajectories of the three-dimensional ODE \eqref{eq:l63pca}, \eqref{eq:gl63}, 
projected on $a_1$ and $a_2$,  
with those of the fitted SDE \eqref{eq:l63pcat}; 
the difference in smoothness of the solutions of ODEs and
SDEs is apparent at that level, although  it can be seen that the fitted SDE model shows a similar pattern of irregular switching between the two distinct components of  the 
attractor in the true system.

Despite the seemingly simple formulation of this example as outlined in section~\ref{sec:PF}, it is a rather difficult problem. On the one hand, the transformed Lorenz 63 system
\eqref{eq:l63pca}, \eqref{eq:gl63} is well known as a classical chaotic system. On the other hand, chaos cannot arise in the reduced-order 2D model of the form \eqref{eq:l63pcat} in a deterministic setting where the noise levels $\sigma_i$ are set to zero, by  the Poincar\'e--Bendixson theorem. \citet{palmer2001nonlinear} showed that a reduced-order 2D model of the Lorenz 63 system 
in discrete time could exhibit \emph{qualitative} features of the original 3D system,
if subjected to additive noise, but it remained an unsolved problem to fit
a stochastic 2D model that \emph{quantitatively} captures detailed chaotic behavior of the original 3D chaotic system. This motivated us to introduce both more unknown parameters and more observed statistics as data, compared to the 
previous simulation study presented in \eqref{sssec:sim}. 
In so doing, we have demonstrated
some success in fitting an SDE model to data from the ODE. We face similar challenges in the remaining examples in this section, where the data are generated from a more complex
model than the model that it is fitted, or indeed is actual observed data.

\subsection{Lorenz 96 System}
\label{ssec:NL96}

We generate data from the slow variable $\{x_k\}_{k=1}^K$ in \eqref{eq:l96}, and use it to fit a
parameterized function $\psi(\cdot)$ and parameter $\sigma$ appearing in \eqref{eq:l96c}. In the
experiments presented below, we take $K=36.$ The form of the data in case (a) is finite-time averaged approximations of $\{\cG_1(x),\cG_2(x)\}$, i.e., observations of the first and second moments of the state vector $x=\{x_k\}_{k=1}^n$. The model discrepancy term $\psi(X_k)$ in \eqref{eq:l96c} is parameterized by the mean of a GP with mean values at $7$ fixed nodes, and constant observation error and hyper-parameters. Thus, we are fitting $10$ parameters for the ODE and $11$ 
for the SDE, using a data vector $y$ of dimension $44$ (when only observing the first $8$ slow variables). The form of the data in case (b) is finite-time averaged approximations of moments up to fourth-order (only evaluating for each single slow variable and thus providing 144 data points) and 11 points on the averaged autocorrelation, leading to a data vector $y$ of dimension $155$. The model discrepancy term $\psi(X_k)$ in \eqref{eq:l96c} is parameterized by the mean of a GP with mean values at $11$ fixed nodes, and constant observation error and hyper-parameters. It should be noted that we also learn $h^2c/J$ in case (b). Thus, we are fitting $15$ parameters for the ODE and $16$ 
for the SDE. The ensemble size is chosen as 300 for case (b). 
Our numerical results illustrate several interesting properties of stochastic
closures for the Lorenz 96 multiscale system:
\begin{enumerate}[(a)]
\item we show that for the relatively large
time scale separation of $c=10$ it is possible to fit both an accurate ODE ($\sigma=0$) and
SDE ($\sigma>0$), in the sense that both ODE and SDE fitted models \eqref{eq:l96} accurately reproduce the  invariant measure of the full system \eqref{eq:l96}, using only $n=8;$ 
\item we show that with weaker scale separation of c = 3, the ODE fit is very poor, but for the SDE it becomes excellent.
\end{enumerate}

\begin{figure}[!htbp]
  \centering
  \includegraphics[width=0.2\textwidth]{G_legend}\\
  \subfloat[ODE]{\includegraphics[width=0.44\textwidth]{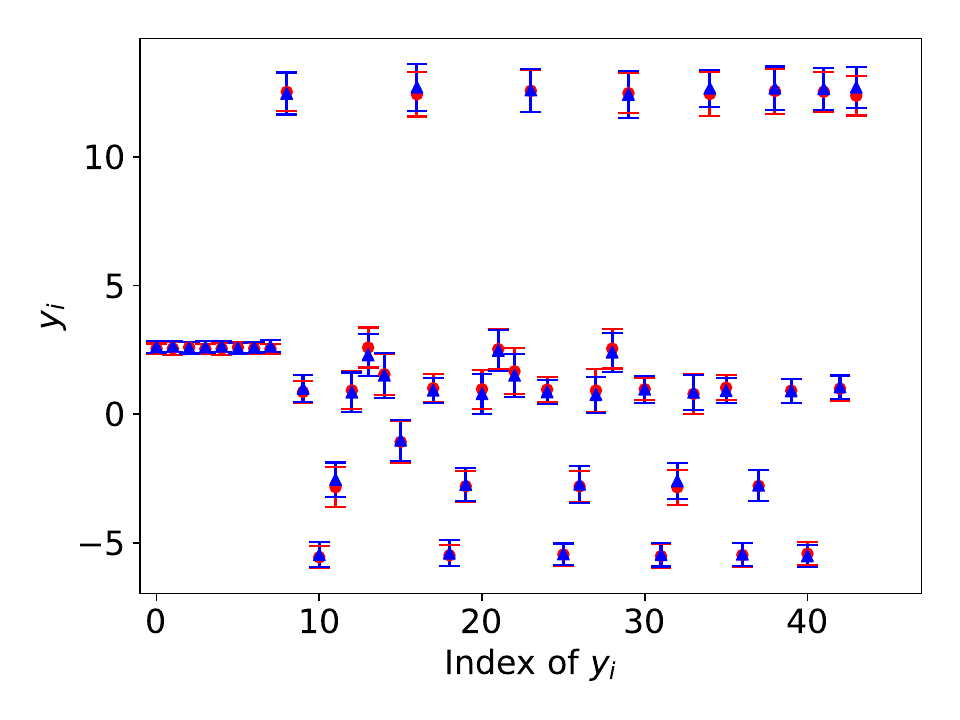}}
  \subfloat[SDE]{\includegraphics[width=0.44\textwidth]{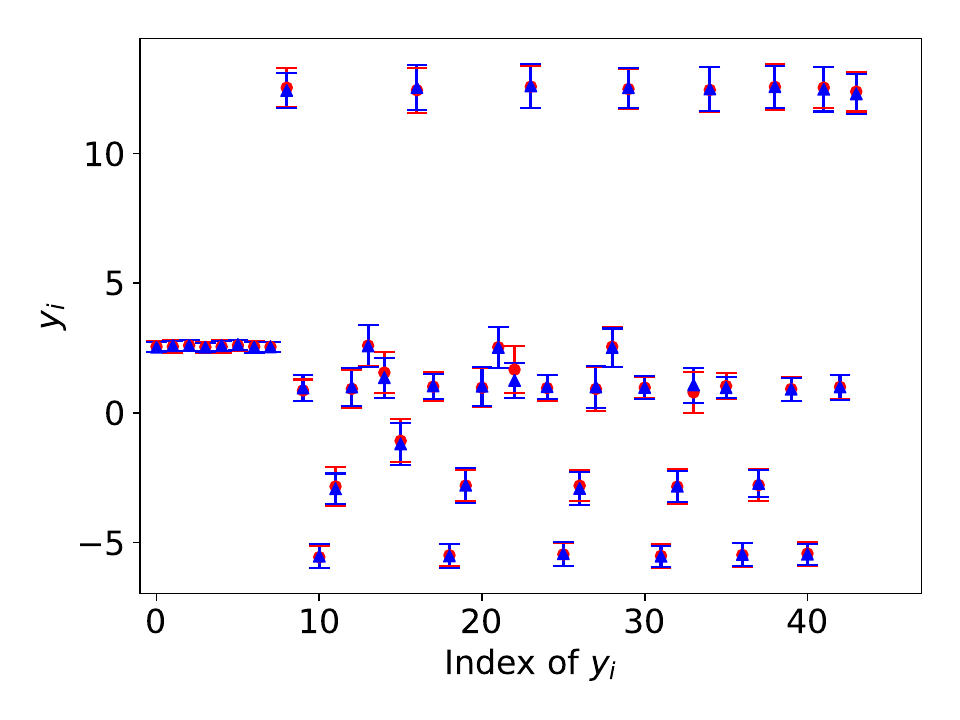}}
    \caption{Comparison of observation data between the true Lorenz 96 system ($c=10$) and the fitted ODE and SDE models.}
  \label{fig:G-L96-c_10-uniform_prior}
\end{figure}

\begin{figure}[!htbp]
  \centering
  \includegraphics[width=0.2\textwidth]{measure_legend}\\
  \subfloat[ODE]{\includegraphics[width=0.44\textwidth]{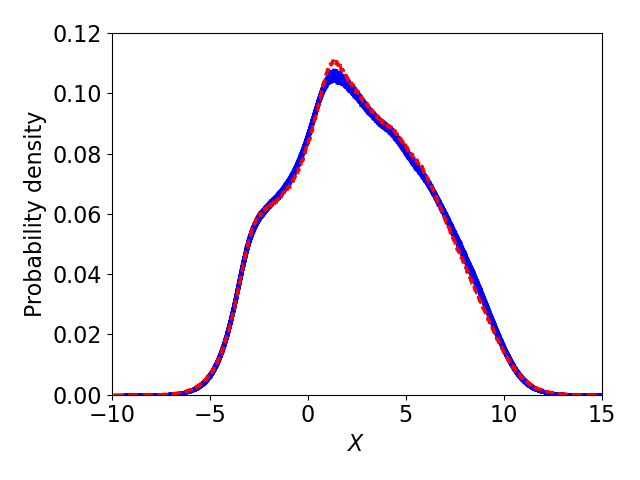}}
  \subfloat[SDE]{\includegraphics[width=0.44\textwidth]{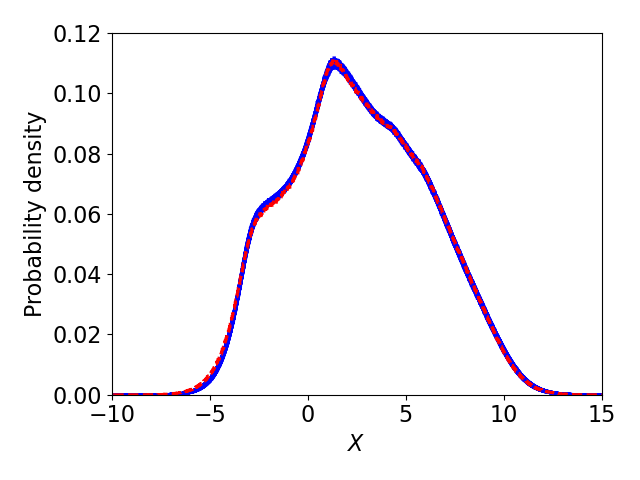}}
    \caption{Comparison of invariant measures between the true Lorenz 96 system ($c=10$) and the fitted ODE and SDE models.}
  \label{fig:Invariant-measure-true-cov-L96-c_10-uniform_prior}
\end{figure}

\begin{figure}[!htbp]
  \centering
  \subfloat[$c=10,~h=1$]{\includegraphics[width=0.44\textwidth]{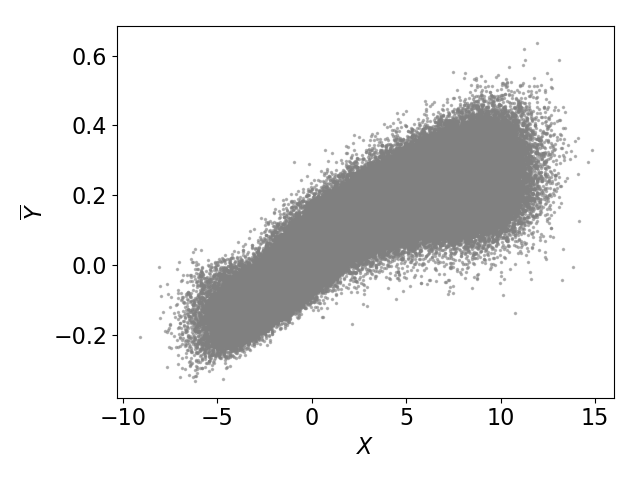}}
  \subfloat[$c=3,~h=10/3$]{\includegraphics[width=0.44\textwidth]{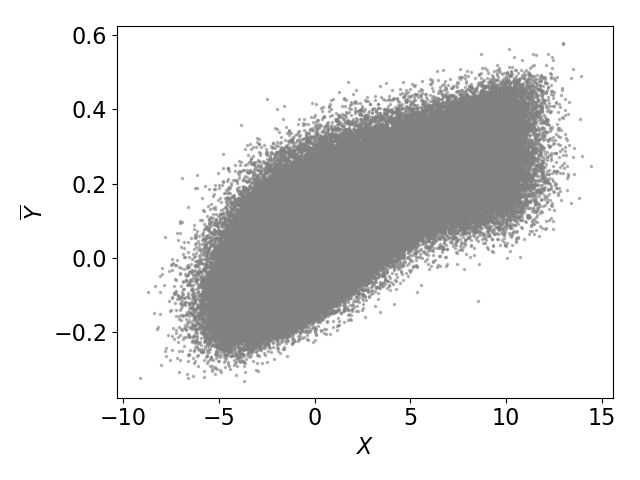}}
    \caption{True closure term of slow variables with different temporal scale separations between slow and fast variables.}
  \label{fig:L96-closure-temporal-scales}
\end{figure}

In case (a), the fixed GP nodes are 7 points uniformly distributed in $[-15,15]$. The initial ensemble of noisily observed values on those nodes are uniformly drawn from $[-1,1]$. The initial ensemble of GP observation error is uniformly drawn from $[0.1,1]$, and the initial ensemble of GP hyper-parameters $a$ and $\ell$ are uniformly drawn from $[0.1,1]$ and $[5,20]$. In
the SDE case, the initial ensemble of $\sqrt{\sigma}$ is drawn from $[0.01,10]$. The trajectory initial condition is uniformly drawn from $[0,1)$ for each state variable. 20 EKI iterations is used. Results for case (a) are presented in Figs~\ref{fig:G-L96-c_10-uniform_prior} and~\ref{fig:Invariant-measure-true-cov-L96-c_10-uniform_prior}. It can be seen in Fig.~\ref{fig:G-L96-c_10-uniform_prior} that both the fitted ODE model and SDE model show almost perfect agreements with the true system in data space. Furthermore, both fitted ODE model and SDE model agree well with the true system when we compare the invariant measure with that of the underlying
data-generating model, in Fig.~\ref{fig:Invariant-measure-true-cov-L96-c_10-uniform_prior}. Although the agreement is slightly better for the fitted SDE model in Fig.~\ref{fig:Invariant-measure-true-cov-L96-c_10-uniform_prior}b, the performances of fitted ODE and SDE models are quantitatively close to each other. The good performance of the fitted ODE model can be explained by invoking the 
averaging hypothesis, as proposed in \cite{fatkullin2004computational},
suggesting a closed ODE model of the form \eqref{eq:l96c}. 
The data implying the form of the closure $\psi$ is as presented in Fig.~\ref{fig:L96-closure-temporal-scales}a. Specifically, the scattering of the true closure term is narrower for $c=10$, indicating that a deterministic closure of slow variables can achieve good agreement with the true system.

\begin{figure}[!htbp]
  \centering
  \includegraphics[width=0.2\textwidth]{G_legend}\\
  \subfloat[ODE]{\includegraphics[width=0.44\textwidth]{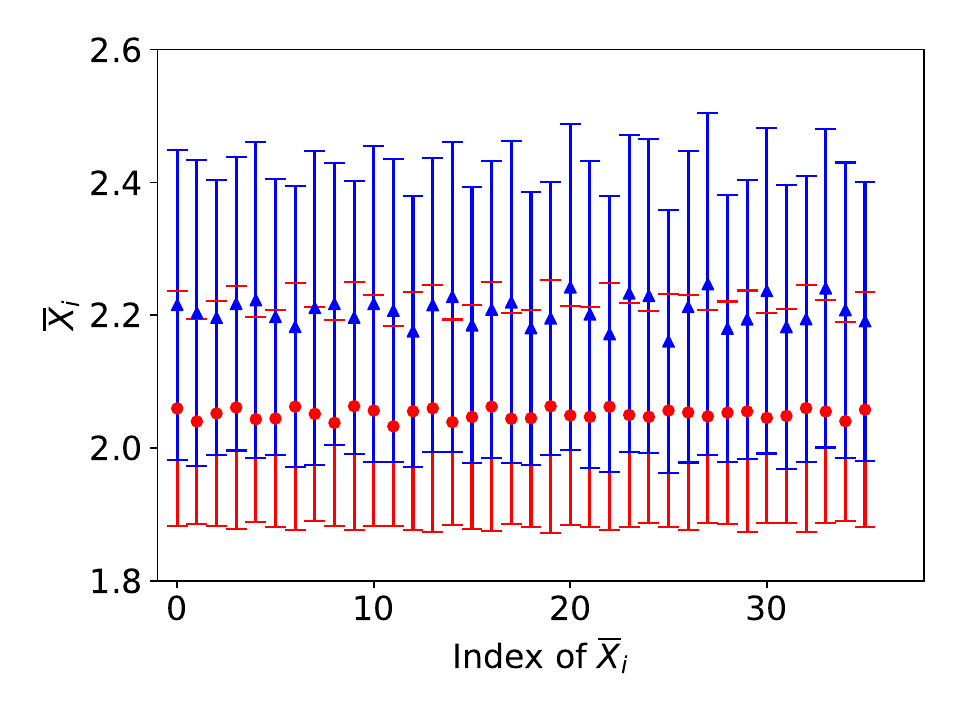}}
  \subfloat[SDE]{\includegraphics[width=0.44\textwidth]{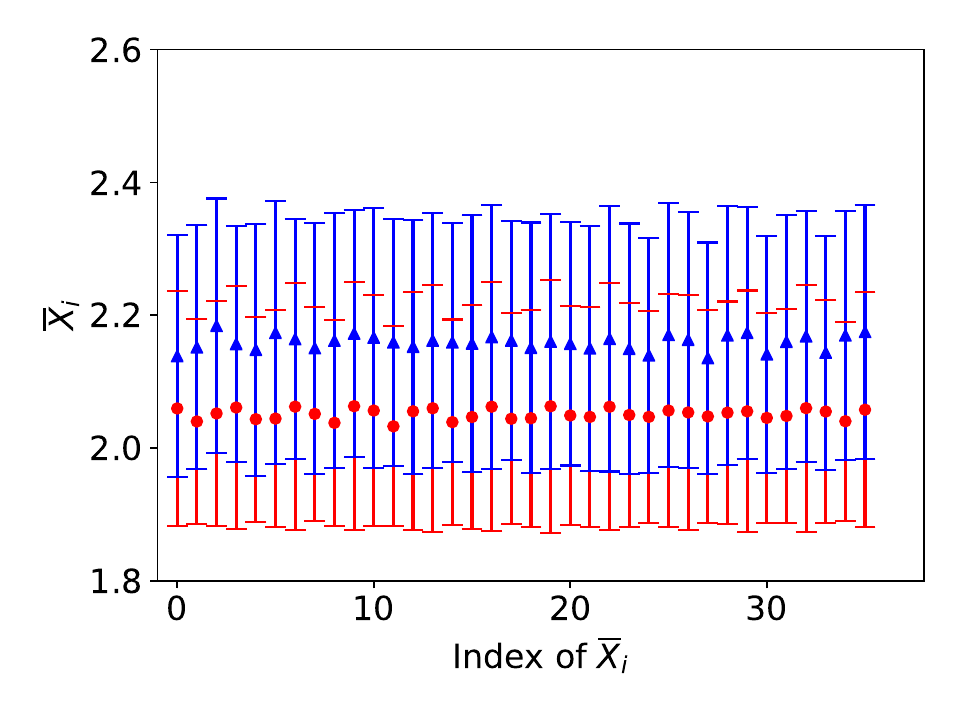}}
    \caption{Comparison of data between the true Lorenz 96 system ($c=3$) and the fitted ODE and SDE models. All 36 slow variables are used to compute the moments as data (only the first moment data are presented here).}
  \label{fig:G-L96-c_3-uniform_prior-full-obs}
\end{figure}

\begin{figure}[!htbp]
  \centering
  \includegraphics[width=0.2\textwidth]{measure_legend}\\
  \subfloat[ODE]{\includegraphics[width=0.44\textwidth]{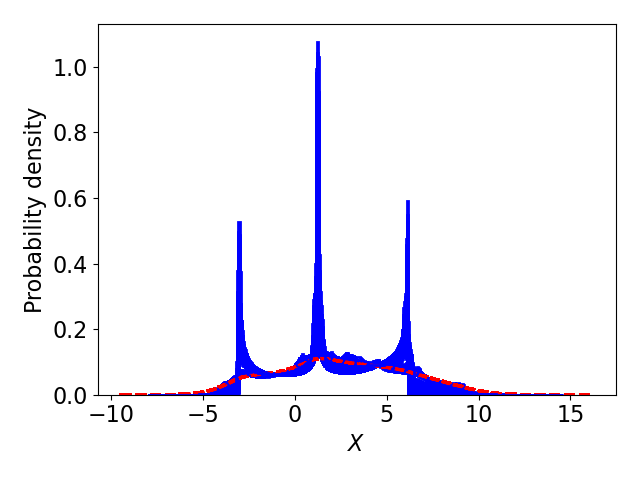}}
  \subfloat[SDE]{\includegraphics[width=0.44\textwidth]{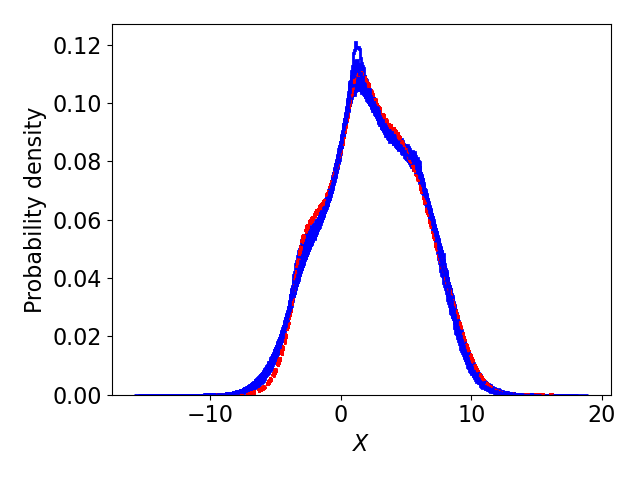}}
    \caption{Comparison of invariant measures between the true Lorenz 96 system ($c=3$) and the fitted ODE and SDE models.}
  \label{fig:Invariant-measure-true-cov-L96-c_3-uniform_prior-full_obs}
\end{figure}

\begin{figure}[!htbp]
  \centering
  \subfloat[ODE]{\includegraphics[width=0.44\textwidth]{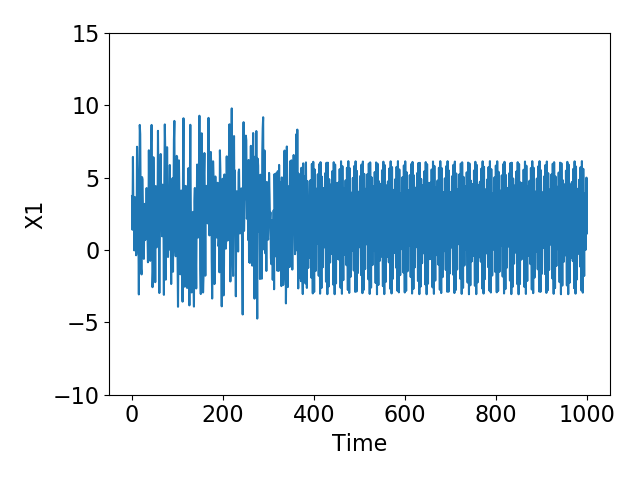}}
  \subfloat[SDE]{\includegraphics[width=0.44\textwidth]{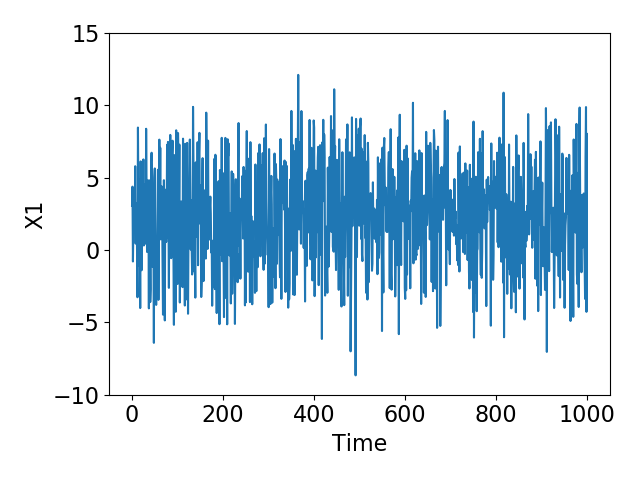}}
    \caption{Comparisons of time series of slow variable $X_1$ by using (a) the fitted ODE model and (b) the fitted SDE model. The time range over which we collect data for EKI is $[0,100]$.}
  \label{fig:X-trajectory-L96}
\end{figure}

In case (b), the fixed GP nodes are 11 points uniformly distributed in $[-15,15]$. The initial ensemble of noisily observed values on those nodes are uniformly drawn from $[-1,1]$. The initial ensemble of GP observation error is uniformly drawn from $[0.1,1]$, and the initial ensemble of GP hyper-parameters $a$ and $\ell$ are uniformly drawn from $[0.1,1]$ and $[1,10]$. The initial ensemble of $h^2c/J$ is uniformly drawn from $[1/3,20/3]$. In the SDE case, the initial ensemble of $\sqrt{\sigma}$ is drawn from $[0.01,10]$. The 
trajectory initial condition is uniformly drawn from $[0,1)$ for each state variable. 20 EKI iterations is used. Results for case (b) are presented in Figs.~\ref{fig:G-L96-c_3-uniform_prior-full-obs} to~\ref{fig:X-trajectory-L96}. The averaging hypothesis
does not apply so cleanly in this case where $c=3$, as can be seen by comparing
the moderate scattering in the data when $c=10$ (Fig.~\ref{fig:L96-closure-temporal-scales}a),  with that obtained when $c=3$ (Fig.~\ref{fig:L96-closure-temporal-scales}b). In the second setting, it turns out that the fitted ODE model is far less satisfactory than the fitted SDE model. Figure~\ref{fig:G-L96-c_3-uniform_prior-full-obs} shows that the SDE fit
achieves similar agreement as the ODE fit in data space. As presented in Fig.~\ref{fig:Invariant-measure-true-cov-L96-c_3-uniform_prior-full_obs}, the fitted ODE model still tends to concentrate toward a small number of discrete values in the long-time behavior, while the invariant measure of fitted SDE model demonstrates excellent agreement with the true system. The issue of the fitted ODE model in failing to maintain correct chaotic behavior for a long time can be more clearly seen in the time series presented in Fig.~\ref{fig:X-trajectory-L96}. More specifically, Figure~\ref{fig:X-trajectory-L96} demonstrates that the fitted ODE model is able to maintain the chaotic behavior within the time range $[0,100]$ that is used to evaluate time-averaged statistics as data. Beyond that time range, the fitted ODE model is
attracted to a quasi-periodic regime, and then stays in that regime afterwards.

\subsection{El Ni\~no–Southern Oscillation (ENSO)}
\label{ssec:NENSO}
\begin{figure}[!htbp]
  \centering
  \includegraphics[width=0.2\textwidth]{G_legend}\\
  \subfloat[Moments]{\includegraphics[width=0.33\textwidth]{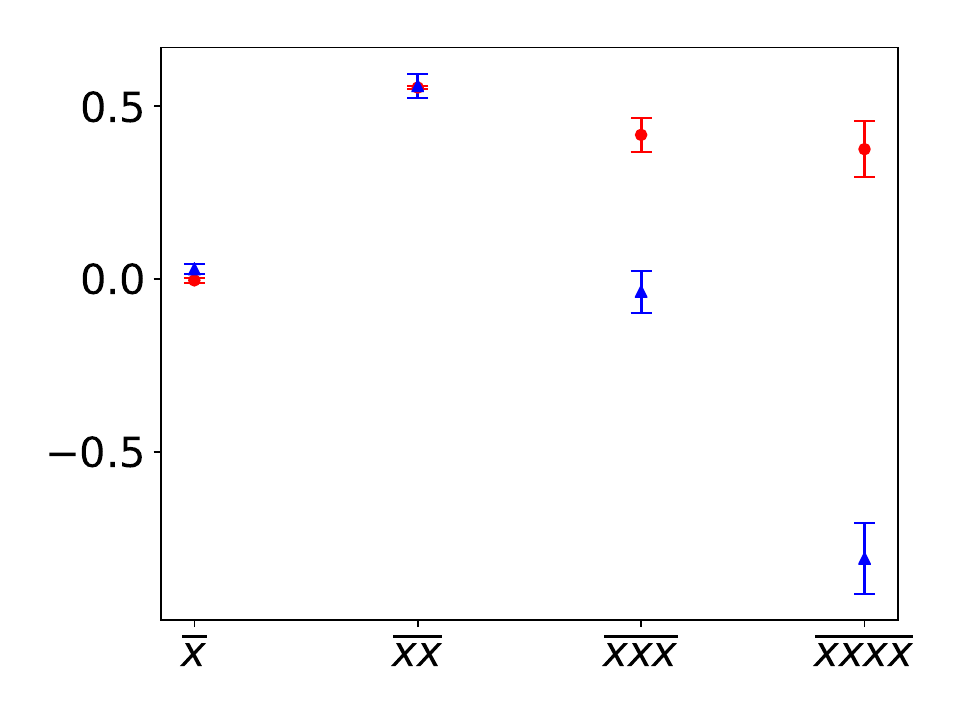}}
  \subfloat[Autocorrelation]{\includegraphics[width=0.33\textwidth]{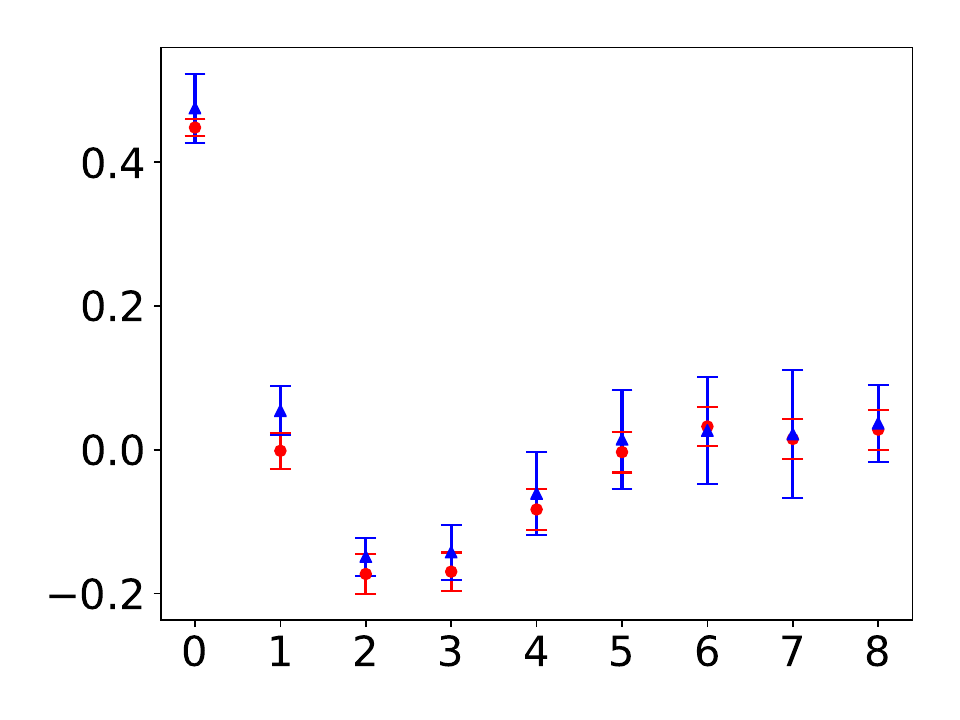}}
  \subfloat[PSD]{\includegraphics[width=0.33\textwidth]{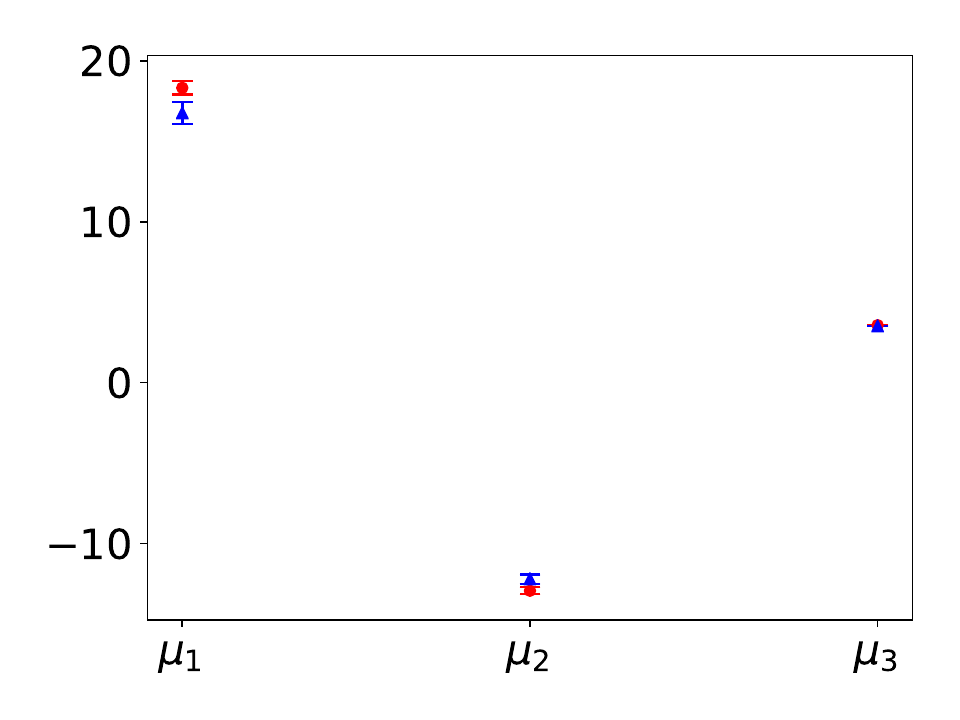}}
    \caption{The comparison of observed quantities, including (a) the first four moments, (b) the autocorrelation function, and (c) the coefficients of fitted second-order polynomial of power spectrum density.}
  \label{fig:ENSO-real-obs-EKI}
\end{figure}

\begin{figure}[!htbp]
  \centering
  \includegraphics[width=0.2\textwidth]{measure_legend}\\
  \includegraphics[width=0.5\textwidth]{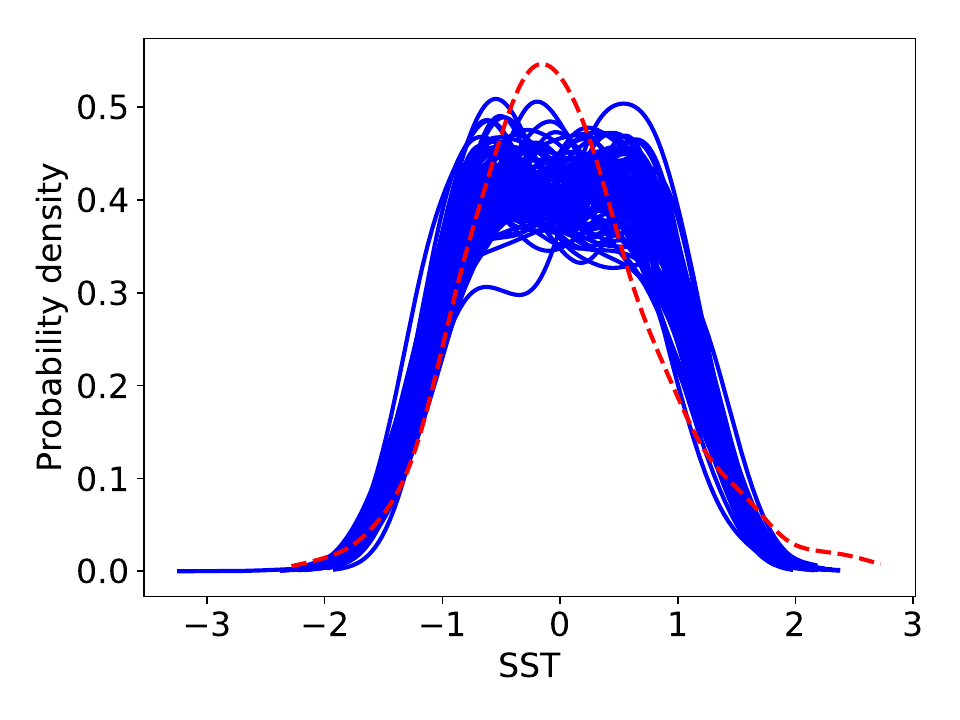}
    \caption{The comparison of the invariant measures of SST between true ENSO data and fitted SDE model.}
  \label{fig:ENSO-real-measure-ensemble}
\end{figure}

In this subsection we use the time-series data \citep{ENSO} for sea surface temperature (SST) $T$ shown in Fig.~\ref{fig:ENSO-data}a to fit an SDDE of the form \eqref{eq:delay-oscillator}, 
learning the four parameters
$a,b,c,\sigma.$ Recall that the delays $\tau_1$ and $\tau_2$ may be viewed as known
properties. The precise form of the data in this case is a finite-time average approximation of $\{\{\cG_{m}(T)\}_{m=1}^4,\cG_{ac}(T),\cG_{psd}(T)\}$. Thus we are fitting 
$4$ parameters in the SDDE by using a data vector $y$ of dimension 16. The initial ensemble of $a,b,c$ are drawn uniformly from $[0.1,10]$. 
The initial ensemble of $\log(\sqrt{\sigma})$ is drawn uniformly from $[\log(0.1),\log(10)]$. The trajectory initial condition is uniformly drawn from $[0,1)$ for each state variable. 10 EKI iterations is used. The true moment data are presented in Fig.~\ref{fig:ENSO-real-obs-EKI}a. To use the autocorrelation function $\cG_{ac}(T)$ as data, we sample $9$ points from it with an interval of six months as presented in Fig.~\ref{fig:ENSO-real-obs-EKI}b. We also use the coefficients of a second-order polynomial fit to the logarithm of the power spectral density; the three coefficients are presented in Fig.~\ref{fig:ENSO-real-obs-EKI}c. Results 
demonstrating the fit are presented in Figs.~\ref{fig:ENSO-real-obs-EKI} to~\ref{fig:ENSO-real-measure-ensemble}.

It can be seen in Fig.~\ref{fig:ENSO-real-obs-EKI} that the fitted SDE model shows a good agreement with true ENSO statistics at first and second order, but not a higher order.
Looking at the invariant measures presented in Fig.~\ref{fig:ENSO-real-measure-ensemble}, we see evidence that the fitted SDE model can capture the long-time behavior of ENSO. The fitted SDE model does not provide a good agreement with data especially in the $4^\textrm{th}$ order moment, indicating the limitation of the current SDE model, which could be addressed by introducing a more sophisticated model.

\subsection{Butane Molecule Dihedral Angle}
\label{ssec:NButane}
\begin{figure}[!htbp]
  \centering
  \includegraphics[width=0.2\textwidth]{G_legend}\\
  \subfloat[Moments]{\includegraphics[width=0.33\textwidth]{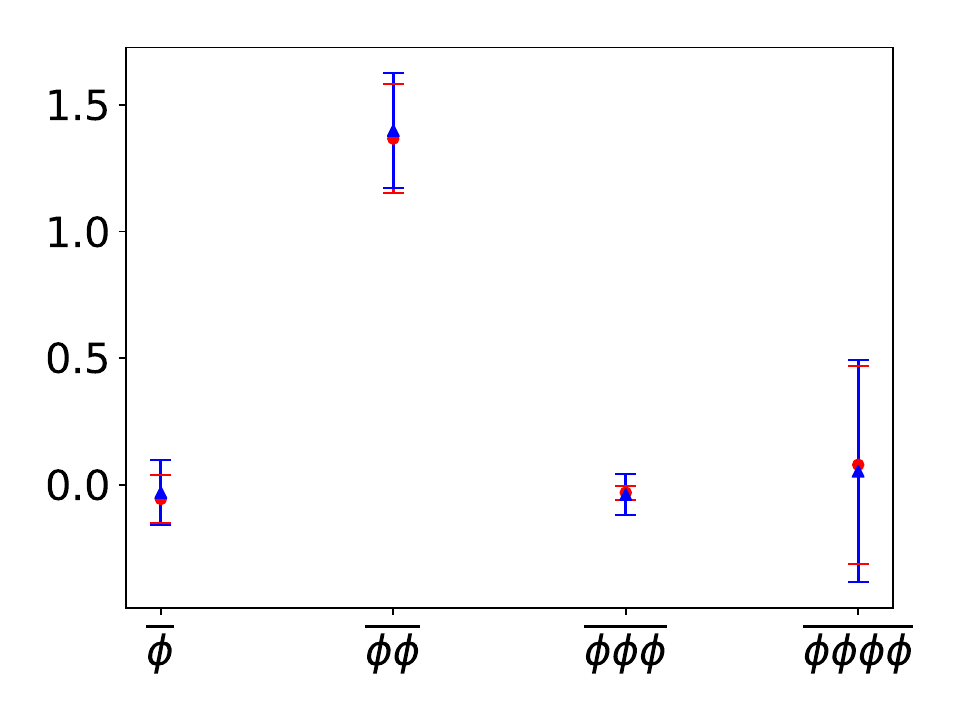}}
  \subfloat[Autocorrelation]{\includegraphics[width=0.33\textwidth]{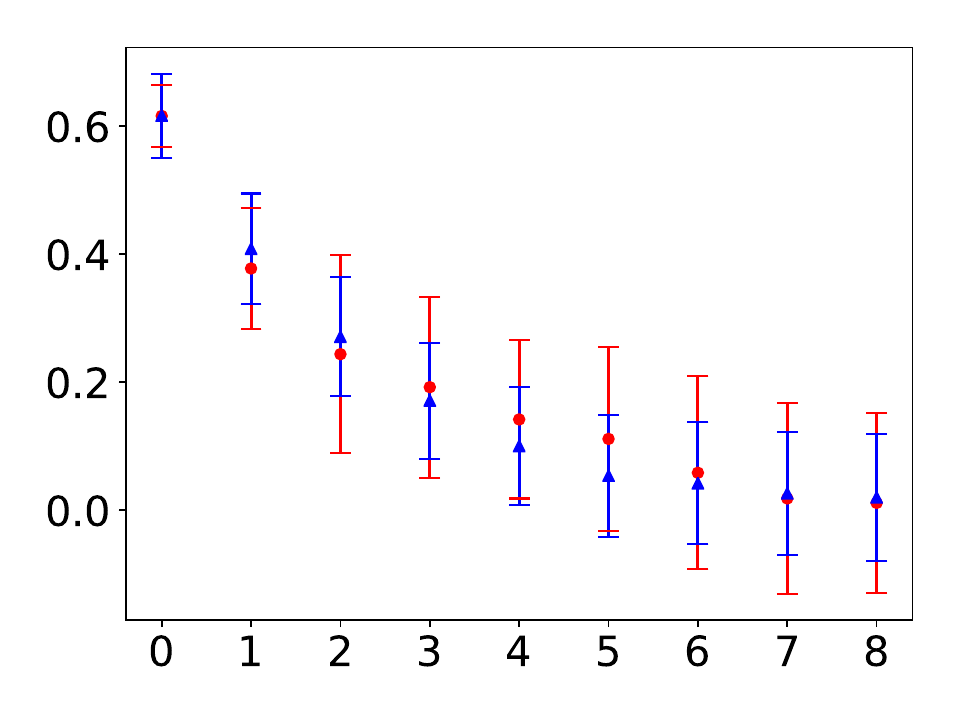}}
  \subfloat[PSD]{\includegraphics[width=0.33\textwidth]{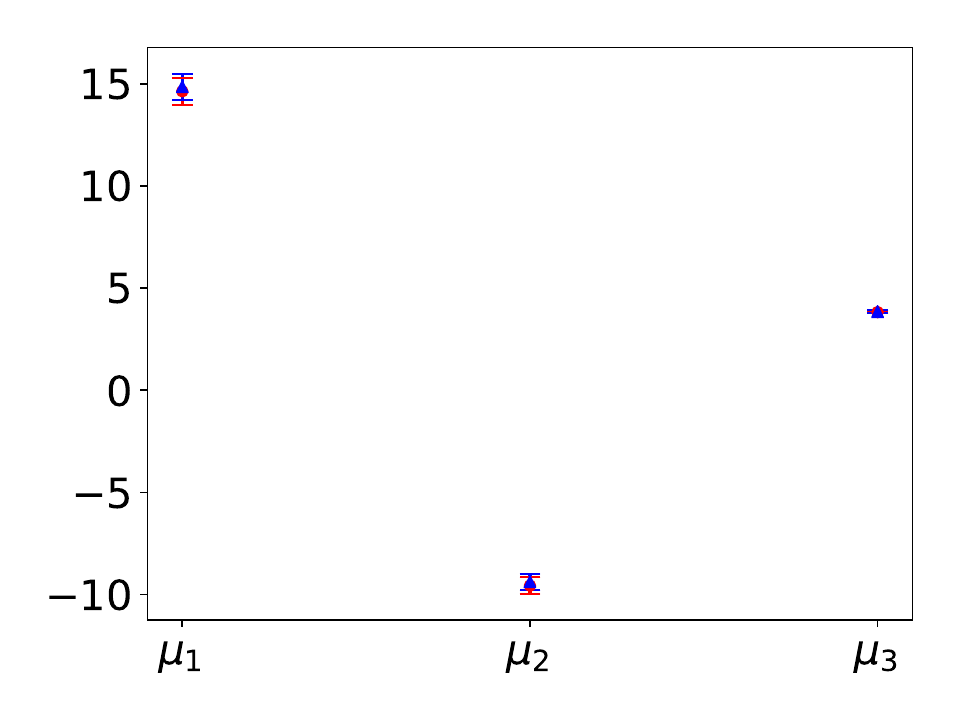}}
    \caption{The comparison of observed quantities of dihedral angle $\phi$, including (a) the first four moments, (b) autocorrelation function and (c) coefficients of fitted second order polynomial of power spectral density.}
  \label{fig:Butane-real-obs-EKI}
\end{figure}

\begin{figure}[!htbp]
  \centering
  \includegraphics[width=0.2\textwidth]{measure_legend}\\
  \includegraphics[width=0.44\textwidth]{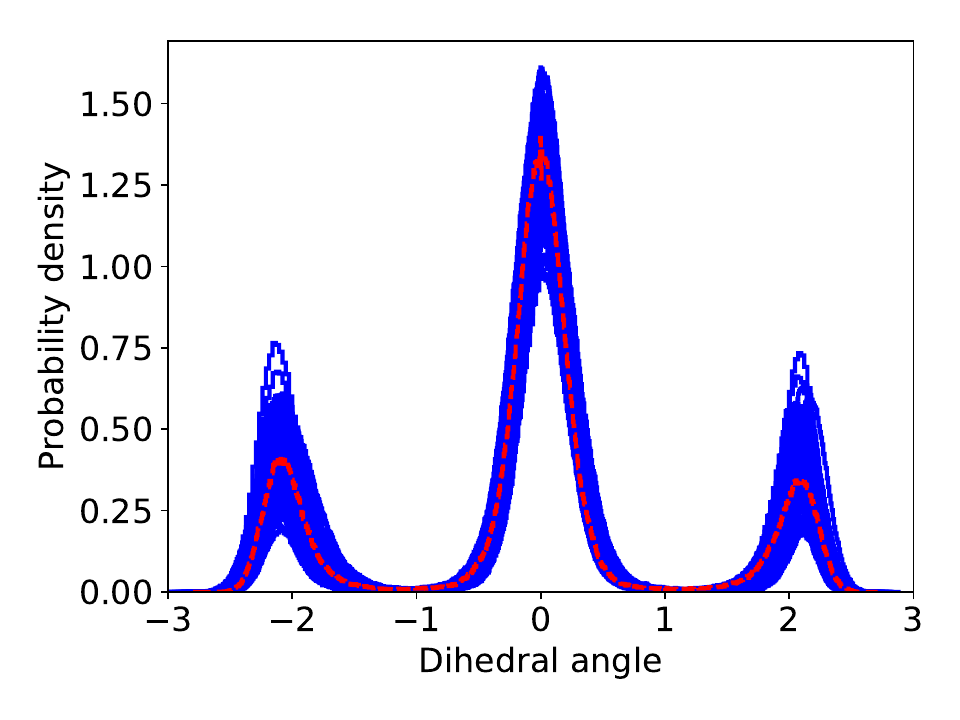}
    \caption{Invariant measures of butane molecule dihedral angle.}
  \label{fig:Invariant-measure-butane}
\end{figure}

\begin{figure}[!htbp]
  \centering
  \subfloat[Truth]{\includegraphics[width=0.49\textwidth]{Butane-timeseries-truth}}
  \subfloat[EKI]{\includegraphics[width=0.49\textwidth]{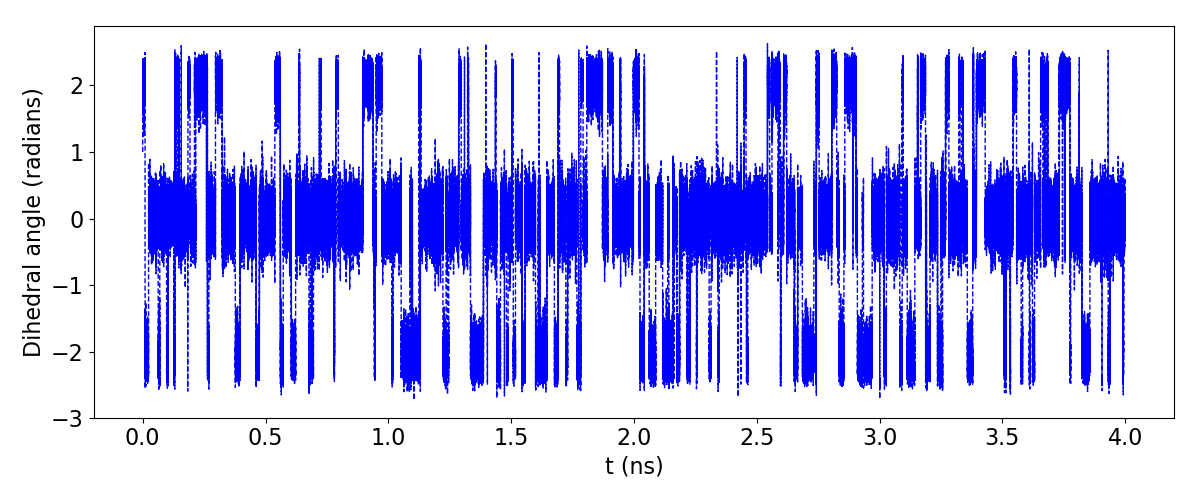}}
    \caption{Time series of butane molecule dihedral angle.}
  \label{fig:time-series-butane}
\end{figure}

\begin{figure}[!htbp]
  \centering
  \subfloat[Truth]{\includegraphics[width=0.49\textwidth]{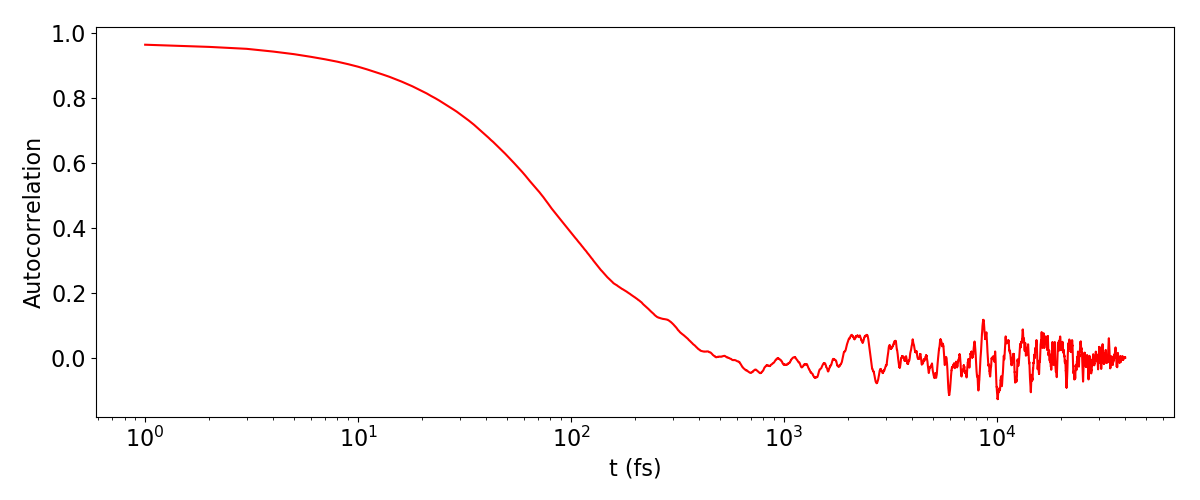}}
  \subfloat[EKI]{\includegraphics[width=0.49\textwidth]{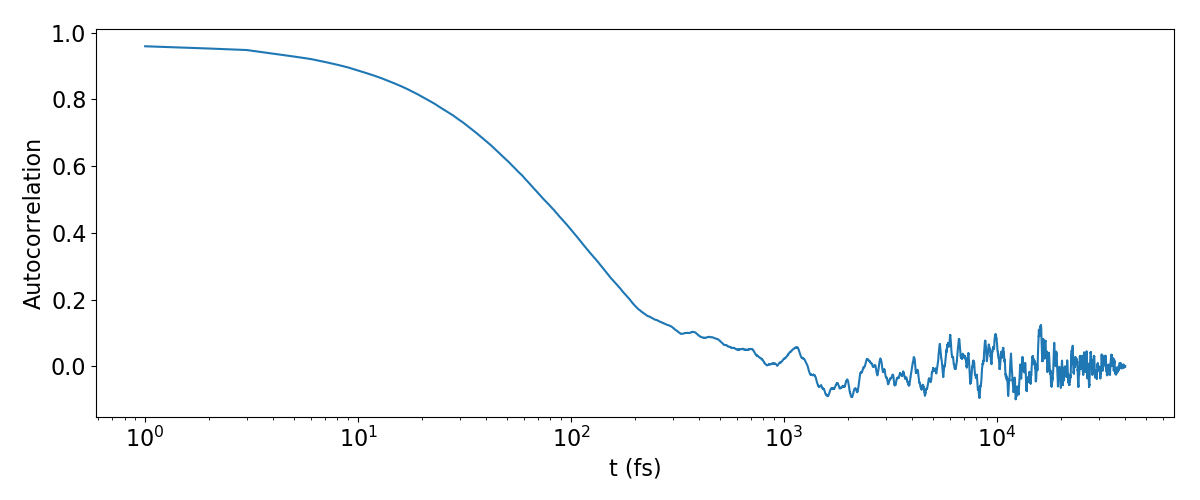}}
    \caption{Autocorrelation of butane molecule dihedral angle.}
  \label{fig:autocorr-butane}
\end{figure}

We fit the second-order Langevin equation \eqref{eq:Langevin2} model to data
derived from \eqref{eq:Langevin1}.
The precise form of the data in this case is a finite-time average approximation of\\ $\{\{\cG_m(\phi)\}_{m=1}^4,\cG_{ac}(\phi),\cG_{psd}(\phi)\}$, where $\phi$ denotes the dihedral angle. Thus, similarly to the previous subsection, we use the first four moments of $\phi$, and the true moment data are presented in Fig.~\ref{fig:Butane-real-obs-EKI}a. Furthermore, $\cG_{ac}(\phi)$ 
is approximated using $9$ points sampled from the autocorrelation function with an interval of $50$ fs (``fs'' stands for femtosencond, and $1~\textrm{fs} = 10^{-15}~\textrm{s}$), and these sampled data points are presented in Fig.~\ref{fig:Butane-real-obs-EKI}b. Finally,
we also use the coefficients of a second-order polynomial that fits to the logarithm of the power spectral density, and the three coefficients are presented in Fig.~\ref{fig:Butane-real-obs-EKI}c. On the other hand, we are learning scalars $\gamma$ and $\sigma$ in \eqref{eq:Langevin1}, together with the potential $\Psi$ constructed from Gaussian basis functions (with length scale fixed as $0.5$) centered at $9$ points evenly distributed in $\left[-\pi,\pi\right]$. Thus, we are fitting $11$ parameters for the SDE by using a data vector $y$ of dimension $16$. The initial ensemble of $\gamma$ and $\log(\sqrt{2\sigma\gamma})$ are drawn uniformly from $[0.1,2]$ and $[\log(0.1),\log(3)]$. The 
trajectory initial condition is uniformly drawn from $[0,1)$ for each state variable. 10 EKI iterations is used. Results showing
the fitted SDE are presented in Figs.~\ref{fig:Butane-real-obs-EKI} to~\ref{fig:autocorr-butane}.

Figure~\ref{fig:Butane-real-obs-EKI} shows that the fitted SDE model achieves very good agreement with all the statistics of true dihedral angle data. More importantly, we can see in Fig.~\ref{fig:Invariant-measure-butane} that the fitted SDE model also leads to an invariant measure that agrees well with that of the true dihedral angle data. 

In order to further validate the fitted SDE model, and in particular to show that it
successfully captures the transition behavior and frequency information of the true data, we further study the time series (Fig.~\ref{fig:time-series-butane}) and autocorrelation function (Fig.~\ref{fig:autocorr-butane}) by simulating the fitted SDE model for a long time.  Figs.~\ref{fig:time-series-butane} and~\ref{fig:autocorr-butane} clearly show that the fitted SDE model successfully reproduces the statistical behavior of the true dihedral angle computed from the much more expensive full molecular dynamics simulation.

\section{Conclusions}
\label{sec:C}

Although computing power has increased dramatically in the past several decades, so too has the
complexity of models that scientists wish to simulate. It is still infeasible to resolve all the interactions within the true system in many applications. SDEs arise naturally as models in many disciplines, even if the first principles governing equations are deterministic, because stochasticity can effectively model unresolved
variables. However, standard statistical methodology for parameter estimation is not always  suitable for fitting SDE models to real data, due to the frequently observed inconsistency between SDEs and real data at small time scales. In this work, we exploit the idea of using sufficient statistics found
by finite-time averaging time series data. Using these statistics, we demonstrated that an SDE model, with the unknown functions being parameterized by Gaussian process regression (GPR), can be fitted to the data using ensemble Kalman 
inversion methods built as optimization-based
variants of the Bayesian sampling algorithms 
proposed in \cite{chen2012ensemble} and \cite{emerick2013ensemble}. 

Ensemble methods are particularly well-suited to this problem
for several reasons: they are derivative-free, thereby sidestepping
computational and technical issues that arise from differentiating
SDEs with respect to parameters; they are inherently parallelizable;
they are robust to the use of approximate, noisy forward model
evaluations; and they scale well to high-dimensional problems. 
Although differentiating SDEs with respect to parameters does
not cause issues in special settings, e.g., when the diffusion coefficient 
is constant so that the stochastic process is not state-dependent, 
differentiation is more problematic when the diffusion coefficient
is state-dependent. Unlike derivative-based optimization methods, ensemble methods avoid differentiating SDEs with respect to parameters and thus are applicable to more general settings. High-dimensional parameter learning would
require regularization in combination with the basic EKI algorithm used here,
as discussed in the introduction. The novel 
hierarchical GPR-based function approximation that we use meshes
well with the EKI methodology.

Future directions of interest in this area include the derivation of
a systematic approach to the determination of sufficient statistics,
analysis of the EKI algorithm for learning in the context of these
problems, and
analysis of the use of GPR-based function representation in nonlinear
inverse problems and the further use of the methodology to new problems
arising in applications. 

\vspace{0.3in}

\noindent{\bf Acknowledgements} The authors thank Yvo Pokern 
(University College London) for providing the butane dihedral 
angle data and giving advice on using it. They are also grateful 
to Sebastian Reich for discussion of several aspects of the
contents of this paper, leading to an improved presentation.
All authors are supported by the generosity of Eric and Wendy Schmidt by recommendation of the Schmidt Futures program, by Earthrise Alliance, Mountain Philanthropies, the Paul G. Allen Family Foundation, and the National Science Foundation (NSF, award AGS1835860). AMS is also supported by NSF (award DMS-1818977) and by the Office of Naval Research (award N00014-17-1-2079).

\end{document}